\documentclass[a4paper,12pt]{article}
\pdfoutput=1
\usepackage{jheppub}

\usepackage[T1]{fontenc} 
\usepackage{amsmath,amssymb,bbm,mathrsfs,nicefrac}
\usepackage{tikz}
\usepackage{xcolor}

\newcommand{\be}{\begin{equation}} 
\newcommand{\ee}{\end{equation}}  
\newcommand{\bea}{\begin{eqnarray}}  
\newcommand{\eea}{\end{eqnarray}}  
\def\bpm{\begin{pmatrix}}
\def\epm{\end{pmatrix}}

\def\gev{\, {\rm GeV}}

\def\bsp#1\esp{\begin{split}#1\end{split}}

\newcommand\lsim{\mathrel{\rlap{\lower4pt\hbox{\hskip1pt$\sim$}}
    \raise1pt\hbox{$<$}}}
\newcommand\gsim{\mathrel{\rlap{\lower4pt\hbox{\hskip1pt$\sim$}}
    \raise1pt\hbox{$>$}}}

\newcommand{\ptcut}{p_T^{\rm cut}}

\newcommand{\captionfonts}{\small}

\newcommand{\approptoinn}[2]{\mathrel{\vcenter{
  \offinterlineskip\halign{\hfil$##$\cr
    #1\propto\cr\noalign{\kern2pt}#1\sim\cr\noalign{\kern-2pt}}}}}

\makeatletter  % Allow the use of @ in command names
\long\def\@akecaption#1#2{%
  \vskip\abovecaptionskip
  \sbox\@tempboxa{{\captionfonts #1: #2}}%
  \ifdim \wd\@tempboxa >\hsize
    {\captionfonts #1: #2\par}
  \else
    \hbox to\hsize{\hfil\box\@tempboxa\hfil}%
  \fi
  \vskip\belowcaptionskip}
  
\makeatother   % Cancel the effect of \makeatletter

\title{Digging for Top Squarks from Higgs data: \\ from signal strengths to differential distributions}

\author[a]{Andrea Banfi,}
\author[a]{Andrew Bond,}
\author[b]{Adam Martin}
\author[a]{and Ver\'onica Sanz}
\affiliation[a]{Department of Physics and Astronomy, University of Sussex, Brighton BN1 9QH, UK}
\affiliation[b]{Department of Physics, University of Notre Dame, Notre Dame, IN 46556, USA}
\abstract{One way to hunt for top squarks is to look for deviations from the Standard Model in loop level processes involving Higgses. This method is indirect, but  complementary to direct searches as it does not rely on specific top squark decays. Studying inclusive Higgs production $pp \to h$ alone is insufficient, since there are parameter regions where the effects of the two top squarks approximately cancel. This degeneracy can be broken by looking at the rate for highly boosted Higgses recoiling against a jet, $pp \to h + \text{jet}$.  In this paper we perform a detailed study of the complementarity of the inclusive and highly boosted processes at the LHC, both in existing Run 1 and Run 2 data, and looking forward to high luminosity. To break the degeneracy, our calculation must maintain the full mass dependence in the loop functions and therefore cannot be recast in an effective field theory framework. We quantify the dependence of both topologies in the top squark parameter space, and outline which levels of experimental and theoretical understanding would be needed for boosted Higgses to be competitive with inclusive Higgs production.
}
\emailAdd{a.banfi@sussex.ac.uk}
\emailAdd{a.bond@sussex.ac.uk}
\emailAdd{amarti41@nd.edu}
\emailAdd{v.sanz@sussex.ac.uk}


\begin{document}
\maketitle
\flushbottom

\section{Introduction}
 
 The second phase of the LHC is well underway and will usher in the era of precision Higgs physics, hunting for any sign of a deviation from the Standard Model (SM) expectations. Being a hadron collider, collisions at the LHC are inherently chaotic and complicated by low-energy QCD effects. This ultimately limits the accuracy at which particle properties can be measured. Future lepton colliders, such as CEPC, FCC-ee, and ILC, CLIC and TLEP~\cite{tlep} offer a cleaner environment and improved precision, but none of the currently discussed possibilities possess the energy of the LHC. 
 
In this paper we study the interplay between the two approaches, precision and energy. As our testing ground, we will use the top squark sector of the minimal supersymmetric standard mode (MSSM). Top squarks modify Higgs properties at loop level~\cite{Kunszt:1991qe,Barger:1991ed,Baer:1991yc,Gunion:1991er,Gunion:1991cw}, and the size and nature of the effects therefore depend on the energies involved. At low energies (on-shell Higgs), top squark loops will modify both Higgs production and decay, which we encapsulate using signal strengths. For our high-energy observable we will look at Higgs plus jet production.  
 
While top squarks can be searched for directly~\cite{stopsearch}, direct searches always rely on assumptions about the decay products, branching ratios, and spectrum. Indirect probes of the top squark sector, such as Higgs coupling measurements, require far fewer assumptions and in most circumstances only depend on the masses and mixings of the top squarks. Effects of top quarks on Higgs production, on the other hand, are independent of the rest of the spectrum and fixed to a large extent by supersymmetry, although related to the sbottom mass spectrum and direct searches~\cite{sbottom}.

Top squark effects in Higgs plus jet production have been explored previously in Ref.~\cite{Grojean:2013nya}, where the authors focused on the region where the stop effects cancel in inclusive production. In Ref.~\cite{Hollik}, the differential rates of Higgs plus jets where discussed in the context of the MSSM, but no explicit expressions or analytical discussion  the differential rates was given. Nowadays, a tool called  {\tt SusHi}~\cite{sushi} allows the user to compute differential rates and interface with Monte Carlo generators such as aMC@NLO~\cite{Mantler:2015vba}. Compared to that work, the present study is more comprehensive, covering a wider array of top squark scenarios and comparing the top squark sensitivity of inclusive Higgs cross section measurements with that of highly boosted Higgses. We also provide a more detailed discussion of the calculation of Higgs plus jet with scalar contributions, with cross checks in the soft and collinear limits.  Our differential rates are computed using a fast Gaussian integrator, allowing the user to obtain the differential rates in a fraction of seconds, hence it is useful when dealing with spectrum parameter scans. There are also several studies of generic new physics loops affecting Higgs production within the framework of effective field theory~\cite{Dawson:2014ora,Ghosh:2014wxa,Grazzini:2015gdl,Dawson:2015gka,Edezhath:2015lga,Grazzini:2016paz} both at leading and next-to-leading (NLO) accuracy, where by definition the mass of the new particles is assumed to be large compared to the momentum of the process. Our study includes the full superpartner mass dependence. As we will show, this additional information allows us to differentiate between supersymmetric spectra that are degenerate when $\sqrt{\hat s} \ll m_{NP}$. The utility of Higgs plus jet has been explored in the past in a similar fashion  to find fermionic top-partners~\cite{Banfi:2013yoa, Azatov:2013hya, Azatov:2013xha, Grojean:2013nya}.

The setup of this paper is as follows: in Sec.~\ref{sec:susy} we define the top squark parameters and explore their contribution to inclusive Higgs production. Next, in Sec.~\ref{sec:H1j}, we introduce Higgs plus jet production and present our analytic results. The full form of the scalar loop contribution to Higgs plus jet is postponed to appendix A along with several cross-checks in special kinematic regions. In Sec.~\ref{sec:mcresult} and Sec.~\ref{sec:comparing_sensitivity}, we introduce the simulation tools and numerically explore the complementarity between inclusive Higgs production and Higgs plus jet for exposing top squark signals.  Finally, in Sec.~\ref{sec:conclude}, we conclude.

%%%%%%%%%%%%%%%%%%%%%%%%%%%%%%%
%%%%%%%%%%%%%%%%%%%%%%%%%%%%%%%
\section{Top squarks and the Higgs}
%%%%%%%%%%%%%%%%%%%%%%%%%%%%%%%
%%%%%%%%%%%%%%%%%%%%%%%%%%%%%%%
\label{sec:susy} 
 
While the minimal supersymmetric Standard Model (MSSM) is a vast framework with rich phenomenology, for the purposes of this work the only aspects of the MSSM that we care about is the top squark couplings to the Higgs and their masses. As such, we are not interested in features that require knowledge of the complete spectrum, such as how the measured Higgs mass is achieved or how/whether there is a viable dark matter candidate.  The top squark sector of the MSSM consists of two complex scalar fields, $\tilde t_L, \tilde t_R$, both of which receive the bulk of their mass from supersymmetry breaking. In addition to $|\tilde t_L|^2,|\tilde t_R|^2$ type masses, the two scalars can mix through interaction with one of the MSSM Higgses. Rather than working with the entries of the top squark mass matrix,  we will parameterize the stop sector by:
\bea
(m_{\tilde{t}_1}  \ , \,  \Delta m \ , \,  \theta),\,
\eea
where the lightest top squark mass is denoted by $m_{\tilde{t}_1}$, the mixing angle by $\theta$ which lies in the interval $[-\pi/2,\pi,2]$, and $\Delta m$ is the separation with the next state, $\Delta m^2=m^2_{\tilde{t}_2}-m^2_{\tilde{t}_1}$. 

With this parametrization, one can write the coupling of the lightest Higgs boson to the mass eigenstate top squarks as
\begin{eqnarray}
\label{eq:ghtt}
g_{h\, \tilde{t}_1 \tilde{t}_1} &=& \frac{m_t^2}{v} \, \left( \alpha_1 c^2_{\theta}+ \alpha_2 s^2_{\theta} + \frac{c_\alpha}{s_{\beta}} \left( 2-\frac{\Delta m^2}{2 m_t^2} s^2_{2 \theta}\right) + \frac{c_{\alpha-\beta}}{s^2_\beta} \frac{\mu}{m_t} s_{2 \theta} \right) \,,\\
g_{h\, \tilde{t}_2 \tilde{t}_2} &=& \frac{m_t^2}{v} \, \left( \alpha_1 s^2_{\theta}+ \alpha_2 c^2_{\theta} + \frac{c_\alpha}{s_{\beta}} \left( 2+\frac{\Delta m^2}{2 m_t^2} s^2_{2 \theta}\right) - \frac{c_{\alpha-\beta}}{s^2_\beta} \frac{\mu}{m_t} s_{2 \theta} \right) \label{eq:ghtt2}\ ,
\end{eqnarray}
where 
\bea
\alpha_1 &=& - \frac{m_Z^2}{m_t^2} \left( 1-\frac{4}{3} s^2_W\right) s_{\alpha+\beta}\,, \\
\alpha_2 &=& - \frac{4}{3} \frac{m_Z^2}{m_t^2}  s^2_W s_{\alpha+\beta}\,, \ 
\eea
and $v= 2 m_W/g \simeq 246$ GeV. Here we use the shorthand notation $s_\beta \equiv \sin \beta,\, c_\beta \equiv \cos \beta, s_W \equiv \sin\theta_W$, etc., where $\tan{\beta}$ is the ratio of the two Higgs vevs and $\alpha$ is the mixing angle rotating the CP-even neutral components of the two Higgses to the mass eigenstates, $h$ and $H$\footnote{In addition to $g_{h \tilde{t}_1 \tilde{t}_1}, g_{h\, \tilde{t}_2 \tilde{t}_2}$ there is a mixed coupling $g_{h\, \tilde{t}_1 \tilde{t}_2}$. We will ignore this coupling throughout since it cannot lead to a 1-loop contribution to $pp \to h$ or $pp \to h + \text{jet}$.}. In the decoupling limit~\cite{needed}, valid as long as the mass of the pseudoscalar $A$ is large compared to the weak scale ($m_A \gg m_Z$), the angles $\alpha$ and $\beta$ are related by $\alpha= \beta -\frac{\pi}{2}$ and the couplings simplify to
\bea
g_{h\, \tilde{t}_1 \tilde{t}_1} &=& \frac{m_t^2}{v} \, \left( \alpha_1 c^2_{\theta}+ \alpha_2 s^2_{\theta} +  2-\frac{\Delta m^2}{2 m_t^2} s^2_{2 \theta}   \right)\,, \\
g_{h\, \tilde{t}_2 \tilde{t}_2} &=& \frac{m_t^2}{v} \, \left( \alpha_1 s^2_{\theta}+ \alpha_2 c^2_{\theta} +   2+\frac{\Delta m^2}{2 m_t^2} s^2_{2 \theta}   \right) \ ,
\eea
where the coefficients $\alpha_{1,2}$ reduce to
\bea
\alpha_1 &=&  \frac{m_Z^2}{m_t^2} c_{2 \beta} \left( 1-\frac{4}{3} s^2_W\right) \,, \\
\alpha_2 &=&  \frac{4}{3} \frac{m_Z^2}{m_t^2} c_{2 \beta} s^2_W  \ .
\eea
We will assume $m_A \gg m_Z$ throughout this paper.

Having defined how the top squarks couple to the lightest Higgs boson, we next study their impact on inclusive Higgs production and Higgs plus jet production. In both cases, the top squarks enter at loop level, induced by gluons and/or quarks. As the stop-gluon coupling is fixed by $SU(3)$  invariance, the top squark contribution to Higgs (and Higgs + jet) production for a given $\hat s$ is a function of the stop masses and mixing alone.

%%%%%%%%%%%%%%%%%%%%%%%%%%%%%%%
%%%%%%%%%%%%%%%%%%%%%%%%%%%%%%% 
\subsection{Top squark contributions to $gg \to h$}
%%%%%%%%%%%%%%%%%%%%%%%%%%%%%%%
%%%%%%%%%%%%%%%%%%%%%%%%%%%%%%%

Focusing first on Higgs production via gluon fusion, the ratio of cross section in the MSSM to the cross section in the SM is given by \cite{Ellis:1975ap,Djouadi:1998az,Khachatryan:2016vau} 
\bea \label{eq:rgludef}
\frac{\sigma_{MSSM}(gg \rightarrow h)}{\sigma^{SM}(gg \rightarrow h)} \simeq \frac{\Gamma(h \rightarrow gg)}{\Gamma^{SM}(h \rightarrow gg)} \equiv \kappa^2_g  \simeq  \left( 1 + \frac{C_g(\alpha_s) \, F_g(m_{\tilde{t}_1},m_{\tilde{t}_2},\theta_{\tilde{t}})}{F_g^{SM}(m_t,m_b \cdots)}  \right)^2\,. \quad \label{kappagdef}
\eea
The function $F_g$ depend on the masses of particles in the loop and their couplings to the Higgs.  For the SM, the only important contribution is the top quark, while in the MSSM loops of both types of top squarks will contribute\footnote{In this comparison we are neglecting subdominant contributions from other squarks.}.  Both the MSSM and SM contributions receive higher order QCD corrections. As these corrections are not the same, there is some residual effect after taking the ratio which we encapsulate into the factor $C_{g}$. Expanding out the mass and coupling dependence of $F_g$ for the cases of interest: 
\bea
 \label{eq:Fgdef}
F_g(m_{\tilde{t}_1},m_{\tilde{t}_2},\theta_{\tilde{t}}) &=&    \sum_{i=\tilde{t}_1,\tilde{t}_2 \cdots} g_{h\,\tilde{t}_i \, \tilde{t}_i}  \, \frac{v}{2 \, m_{i}^2} \, F_0 (\tau_i)\,, \label{F2}  \\
F_g^{SM}(m_t,m_b \cdots) &=&  \sum_{i=t,b \cdots} F_{1/2}(\tau_i)\, \left(1 + \frac{11 \, \alpha_s}{4\, \pi}\right)\,.
\eea
For a given particle $i$ running around the gluon fusion loop, the functions $F_0$ and $F_{1/2}$ depend on the dimensionless variable $\tau_i = m_h^2/(4 m_{i}^2)$  and can be further decomposed as:
\begin{align}
& F_0(\tau)=\left[\tau -f(\tau) \right]/\tau^2,\quad F_{1/2}(\tau)= - 2 {\left[\tau + (\tau -1) f(\tau) \right]}/{\tau^2}\,,\quad  \nonumber \\
& \text{with}\quad f(\tau) = \left\{\begin{array}{cc} \arcsin^2 \sqrt{\tau} & \tau \leq 1 \\ -\frac{1}{4} \, \left[\log \frac{1 + \sqrt{1 - \tau^{-1}}}{1 - \sqrt{1 - \tau^{-1}}} - i \, \pi \right]^2 & \tau > 1 \end{array} \right. \,.
 \end{align} 
 
To gain some insight into Eq.~\eqref{eq:Fgdef}, it is useful to take some limits. If both the top squarks are heavy, $m_{\tilde t_{1,2}} \gg m_h$, the top squark contribution can be matched onto local operators in the context of an EFT analyses~\cite{EFTstuff}, and the functions $F_g, C_g$ simplify. Neglecting the effects of gluinos, squark mixing, and any running that would sum the large logs that would appear if the matching was done in two stages (e.g. one top squark eigenstate at a time), the perturbative matching correction is~\cite{Dawson:1996xz}:
\begin{equation}
C_g(\alpha_s) = 1 + \frac{25 \, \alpha_s}{6 \, \pi}, \\ 
\end{equation}
%Inserted into Eq.~\eqref{eq:rgludef}, and plugging in numerical values for all parameters other than the top squark masses and mixing, $\alpha_s=\alpha_s(m_h/2) = 0.1276$, $m_t$ = 173.5 GeV, $m_b$ = 4.65 GeV and $m_h =  125.26 \pm 0.21$ GeV~\cite{MoriondHiggs}:
%\begin{align}
%\kappa_g  \simeq 1 + (-0.815+0.081i) \, F_g(m_{\tilde{t}_1},m_{\tilde{t}_2},\theta_{\tilde{t}}).
%\end{align}
%
Carrying out $m_{\tilde t_{1,2}} \gg m_h$ in $F_g (m_{\tilde{t}_1},m_{\tilde{t}_2},\theta_{\tilde{t}})$, we find
\bea
 F_g (m_{\tilde{t}_1},m_{\tilde{t}_2},\theta_{\tilde{t}}) &=& - \frac 1 3 \sum\limits_{i=\tilde{t}_1,\tilde{t}_2} \frac{g_{h\,\tilde{t}_i \, \tilde{t}_i}\, v}{2 \, m_{i}^2} \nonumber \\
 &=& -\frac{1}{3} \, \left[\frac{m_t^2}{m_{\tilde{t}_1}^2} + \frac{m_t^2}{m_{\tilde{t}_2}^2}  - \frac{1}{4} \, \sin^2 (2 \, \theta) \, \frac{\Delta m^4}{m_{\tilde{t}_1}^2 \, m_{\tilde{t}_2}^2}\right] \,. \label{expF}
\label{eq:Fg}
\eea
up to corrections of $\mathcal O(g^2)$. Clearly, Eq.~\eqref{eq:Fg} is the sum of two types of terms, a positive-definite contribution from both mass eigenstates, and another dependent on the mixing angle. As such,  the size  -- and even overall sign -- of $ F_g (m_{\tilde{t}_1},m_{\tilde{t}_2},\theta_{\tilde{t}}) $ depends on the details of the mixing in the stop sector. The stop contribution depends on the mass and {\it chirality}  of the eigenstates. If the overall contribution is dominated by a light eigenstate of pure handedness  ($\theta \to 0$) the sum will be negative, whereas if the mixing term proportional to $\Delta m^2$ is dominant then the sum will be positive. Stated another way, in the case of zero mixing we expect there to be an enhancement of $\sigma(g g \to h )$, but in the case where there is sizable mixing and the $\Delta m^2$ term dominates, the separation between the two eigenstates will govern the suppression.  Dialling the mixing between the $\theta \to 0$ limit and the limit of large $\Delta m^2$, it is clear that there are slices of parameter space where  the contributions from the two states partially cancel each other and $ F_g (m_{\tilde{t}_1},m_{\tilde{t}_2},\theta_{\tilde{t}}) \sim 0$.  For these parameter regions, $gg \to h$ will have reduced sensitivity of the top squark sector. We emphasize that, while the possibility of cancellation in $F_g$ when there is large mixing among the top squarks is easiest to see analytically in the $m_{\tilde t_1}, m_{\tilde t_2} \gg m_t$ limit, it is not restricted to that parameter region.

 As we will show below, by adding an extra jet to the final state $pp \to h +\text{jet}$ and focusing on regions where the Higgs has high $p_T$, we can disrupt the cancellation among top squark loops. Whereas these cancellations may be accidental, and hence unmotivated from a model building point of view, adding differential information from $pp \to h +\text{jet}$  can only increase the amount of information we have on the stop sector. Moreover, future measurements of deviations in both total rates and differential rates may allow us to distinguish among different types of stop sectors, pointing out to specific UV realizations of supersymmetry.  

%%%%%%%%%%%%%%%%%%%%%%%%%%%%%% 
%%%%%%%%%%%%%%%%%%%%%%%%%%%%%% 
\subsection{Top squark contributions to $pp \to h + \text{jet}$}
%%%%%%%%%%%%%%%%%%%%%%%%%%%%%% 
%%%%%%%%%%%%%%%%%%%%%%%%%%%%%% 
\label{sec:H1j}

Higgs plus jet production in the SM is a one-loop process induced by $gg, qg, \bar q g$ or $\bar q q$ partons. The top squark contribution has been calculated previously in Ref.~\cite{Hollik,sushi,Grojean:2013nya}; however this is the first instance where the analytic form of the amplitude is given. The analytic form is useful as it allows us to understand how the contributions behave in different kinematic limits. In addition to the analytic expression for the top squark contributions, we also show their behaviour in the soft and collinear limits in appendix~\ref{sec:H1j-soft-coll}. While not strictly required for our numerical analysis, the soft and collinear limits serve as a valuable cross-check.

The largest component of $pp \to h + j$ comes from the $gg$ initial state. The $gg \to hg$ amplitude can be expressed in terms of eight
primitive helicity amplitudes $\mathcal{M}_{h_1 h_2 h_3}$
corresponding to the possible choices for each gluon helicity
$h_i=\pm$. We use the convention that the momenta of gluons
$p_1$ and $p_2$ are incoming, and that of gluon $p_3$ is outgoing, so that the
Mandelstam variables are defined as
\begin{align}
s = (p_1 + p_2)^2\,, \quad
t =(p_2 - p_3)^2\,, \quad
 u = (p_1 - p_3)^2\,. 
\end{align}
The helicity amplitudes are then related to the full, un-averaged
amplitude squared via
\begin{align}
  | M_{gg\to Hg}|^2 &= \frac{N_c (N_c^2 - 1) \alpha_s^3}{64 \pi v^2}
  \sum_{h_1,h_2,h_3 = \pm} \left| \sum_{i=t,b,\tilde{t}_1,\tilde{t}_2}
    \mathcal M^i_{h_1 h_2 h_3} \right|^2 \,.
\label{eq:squaredamp}
\end{align}
The index $i$ here refers to the particle running in the loop needed
to couple the gluons to the Higgs. After applying parity and crossing
symmetry, only two of the helicity amplitudes are independent, which
we take to be $\mathcal M^{i}_{+++}$ and $\mathcal M^{i}_{++-}$. The
amplitudes for fermions in the loops (needed for $i = t,b$) can be
found in appendix A of Ref.~\cite{Glover-Baur}. 

The contributions to the helicity amplitudes due to loops containing a top squark with mass $m = m_{\tilde{t}_i}$  and coupling to the Higgs $g_{h \tilde{t}_i \tilde{t}_i}$, are:
\begin{align}
\mathcal M^{\tilde{t}_i}_{+++} &= (g_{h\, \tilde{t}_i \tilde{t}_i}\, v\, \Delta)\, \times \, \Big\{16\Big(\frac 1{t\,u} + \frac 1{t_1\,t} + \frac 1 {u_1\, u} \Big) + \frac{16}{s} \Big( \frac{B_1(t)(2s + u)}{t^2_1} + \frac{B_1(u)(2s + t)}{u^2_1} \Big) \nonumber \\
& +32\, m^2\Big(\frac{C_1(t)}{t\,t_1} + \frac{C_1(u)}{u\, u_1} \Big) - \frac{16\,m^2}{s\,t\,u}\Big( s_1\, C_1(s) + (u - s)\,C_1(t) + (t-s)\,C_1(u) \Big) \nonumber \\
& + \frac{8\, m^2}{s\,t\,u}\Big( s\,t\,D_0(s,t) + s\,u\,D_0(s,u) - t\,u\,D_0(t,u) \Big) - \frac{16\, m^2}{s}D_0(t,u) + \frac{8}{s^2}\,E_0(t,u)\Big\} \, .
\label{eq:mppp}
\end{align}
and
\begin{align}
\mathcal M^{\tilde{t}_i}_{++-} &= (g_{h\, \tilde{t}_i \tilde{t}_i}\, v\, \Delta)\, \times \Big\{-\frac{16\, m^2_H}{s\,t\,u} + \frac{16\, m^2}{s\,t\,u}\Big( s_1\, C_1(s) + t_1\, C_1(t) + u_1\, C_1(u) \Big) \nonumber \\
& ~~~~~~~~~-\frac{8\,m^2}{s\,t\,u} \Big( s\,t\, D_0(s,t) + s\,u\,D_0(s,u) + t\,u\, D_0(t,u) \Big) \Big\}\,.
\label{eq:mppm}
\end{align}
In these expressions we define
\begin{align}
s_1 \equiv s - m^2_H\,,\quad 
t_1 \equiv t - m^2_H\,,\quad
 u_1 \equiv u - m^2_H\,,\qquad
\Delta = \sqrt{\frac{s\,t\,u}{8}}\,. \label{deltadef}
\end{align}
The functions $B_1, C_1, D_0$ are 1-loop basic scalar integrals. They
are functions of $s,t,u$, the mass of the particle in the loop, and
the Higgs mass; their definitions can be found in
\cite{Passarino-Veltman}. The function $E_0$ introduced in
\cite{Glover-Baur} is an auxiliary function defined as
\begin{equation}
  \label{eq:E0-definition}
  E_0(s,t) = s\, C_0(s) + t\, C_0(t) + s_1\, C_1(s) + t_1\, C_1(t) - s\, t\, D_0(s,t) \, ,
\end{equation}
where $C_0$ is again a 1-loop scalar integral defined in
\cite{Passarino-Veltman}.

The other $pp \to hj$ subprocesses $(q\bar q \to h g,\, qg \to
hq,\, \bar q g \to h \bar q)$ are controlled by a third function,
the un-averaged amplitude squared
\begin{align}
\sum | M_{q\bar{q}\to Hg}|^2(s,t,u)  &= \frac{16 \alpha_s^3}{\pi v^2 s}\frac{t^2 + u^2}{s_1^2} \left| \sum_{i= t,b,\tilde{t}_1,\tilde{t}_2} \mathcal M^i(q \bar{q} \to h g) \right|^2 \,.
\label{eq:squaredampquarks}
\end{align}
The SM amplitudes can be found are in
Ref.~\cite{Glover-Baur}, while for scalars running in the loop we have:
\begin{align}
  \mathcal M^{\tilde{t}_i}(q\bar q \to h g) &= -(g_{h\, \tilde{t}_i \tilde{t}_i}\, v)\times
  \Big( \frac 1 2 + m^2\, C_1(s) + \frac{s}{2\,s_1}\,B_1(s) \Big)\,.
\label{eq:qqbar}
\end{align}
We can get the amplitudes for the subprocesses $qg \to hq$ and $ gq
\to h q$ from the above by swapping the Mandelstam variable
$s$ and $u$, and $s$ and $t$ respectively.

Before we investigate the numerical impact of these corrections\footnote{In our calculations we will assume the only scalars running in the loop are $\tilde{t}_{1,2}$. If
$\tan{\beta}$ is large, the bottom squark loops may be non-negligible. One could account them by
adding the appropriate terms to the sums in
equations~\eqref{eq:squaredamp} and~\eqref{eq:squaredampquarks}. Additionally, there are extra diagrams that must be included coming
from loops involving gluinos. When gluinos decouple, their effect gets
absorbed into a correction of relative order $\alpha_s$ to the
couplings of quarks and squarks to the Higgs. We neglect such
contribution in our calculation of the above matrix elements, since
these effects are of the same order as unknown higher-order QCD
corrections~\cite{Muhlleitner:2008yw}. } to the
$p_T$-spectrum of a Higgs boson recoiling against one jet, let us discuss
the qualitative feature of the above amplitudes. For the sake of
illustration, we consider only $\mathcal{M}_{+++}$, but similar
results hold for other amplitudes as well and are presented in appendix~\ref{sec:H1j-decoupling}.  We are particularly interested in how the SM (found in eq.~(A.15) of ref.~\cite{Glover-Baur}) and top squark contributions behave when the momentum flowing through the loop is either much smaller or much larger than the loop particle masses. We denote $m$ as the scale of the superpartner masses, $m\sim$ $m_{\tilde t_{1,2}}$, and distinguish two regimes:
\begin{itemize}
\item Low-$p_T$ limit $p_T \ll m$. Using the results in
  eq.~(\ref{eq:M-decouple}), we find that the scalar contribution reduces to 
  \begin{equation}
    \label{eq:Mppp-stop-lowpt}
    \left.\mathcal{M}_{+++}^{\tilde t_i}\right|_{p_T\ll m} \simeq -\frac{4}{3}\frac{\Delta}{p_T^2} \frac{g_{h\, \tilde{t}_i \tilde{t}_i}\, v}{m^2}\sim p_T \frac{m_t^2}{m^2}\,,
  \end{equation}
  where we have used the definition of $\Delta$~\cite{Glover-Baur}, Eq.~\eqref{deltadef}, and the fact that $\Delta \sim p_T^3$. The factor $1/m^2$ originates from triangle loops. Finally, we have used the fact that the leading soft SUSY breaking part of the coupling $g_{h\, \tilde{t}_i \tilde{t}_i}$ is  proportional to $m_t^2/v$, see Eq.~\eqref{eq:ghtt2}.
  In the SM case, the triangle loops contribute a factor of
  $1/m_t^2$ but have two extra powers of $m_t$ in
  the numerator  -- one from the Yukawa coupling of the top, the other
  from a helicity flip imposed by the interaction with the Higgs. As a result:
\begin{equation}
    \label{eq:Mppp-top-lowpt}
    \left.\mathcal{M}_{+++}^{t}\right|_{p_T\ll m_t} \simeq -\frac{32}{3}\frac{\Delta}{p_T^2}\sim p_T \,,
  \end{equation}
  with no dependence on the top mass, as in the total cross section
  $gg\to h$. Note that this low $p_T$ limit holds with a very good
  approximation also in the region $p_T \sim m,m_t$.
\item High-$p_T$ limit $p_T \gg m,m_t,m_H$. In this limit, the amplitudes for both squarks and fermion loops reduce to single and double logarithms of
  $p_T/m$, 
\begin{subequations}
    \begin{align}
      \left.\mathcal{M}_{+++}^{t}\right|_{p_T\gg m_t}& \simeq \frac{m_t^2}{p_T}
      \left(A_0+A_1\ln\left(\frac{p_T^2}{m_t^2}\right)+A_2\ln^2\left(\frac{p_T^2}{m_t^2}\right)\right) \,,\label{eq:hipt} \\
      \left.\mathcal{M}_{+++}^{\tilde t_i}\right|_{p_T\gg m}& \simeq \frac{g_{h\, \tilde{t}_i \tilde{t}_i}\, v}{p_T}\left(\tilde A_0+\tilde A_1\ln\left(\frac{p_T^2}{m^2}\right)+\tilde A_2\ln^2\left(\frac{p_T^2}{m^2}\right)\right)\,,
    \end{align}
  \end{subequations}
  where $A_i,\tilde A_i\>(i=0,1,2)$ do not depend on the mass of the
  particle in the loops, but only on kinematic invariants. Also, $A_i$
  are the same for all processes involving any fermion in the loops
  coupling in the same way a top does (e.g. a top partner), while
  $\tilde A_i$ are the same for all process with scalars in the loops.
\end{itemize}

Having seen the behavior of different components of the Higgs plus jet amplitude in the low and high-$p_T$ regime, we can combine things to get a sense of the behavior of the amplitude as a whole. In the limit where the $p_T$ of the Higgs is much less than $m_t$ or either of the top squark masses, the combination of Eq.~\eqref{eq:Mppp-top-lowpt}, \eqref{eq:Mppp-stop-lowpt} and \eqref{eq:Fg} 
yields:
\begin{equation}
  \label{eq:M+++-lowpt}
  \left.\mathcal{M}_{+++}\right|_{p_T\ll m_t,m_{\tilde t_1},m_{\tilde t_2}} \simeq -\frac{32}{3}\frac{\Delta}{p_T^2} - \frac{4}{3}\frac{\Delta}{p_T^2} \sum\limits_{i = \tilde t_1, \tilde t_2} \frac{g_{h\, \tilde{t}_i \tilde{t}_i}\, v}{m_{\tilde t_i}^2}  \simeq 8 \frac{\Delta}{p_T^2}\left(-\frac{4}{3}+ F_g (m_{\tilde{t}_1},m_{\tilde{t}_2},\theta_{\tilde{t}})\right).
\end{equation}
In this limit, the top squark contributions are combined into $F_g$, the same function appearing in inclusive production. In fact, in the $m_t \to \infty, m_{\tilde t_1}, m_{\tilde t_2} \gg m_t$ limit, 
\bea
\label{kappaap}
\kappa_g \to (1-\frac 3 4 F_g) ,
\eea
the exact combination appearing in Eq.~\eqref{eq:M+++-lowpt}. As such, the pattern of deviations in the low-$p_T$ regime of $h + \text{jet}$ production will mirror those of $pp \to h$. In particular, the parameter regions where the two stops cancel (e.g. when $F_g (m_{\tilde{t}_1},m_{\tilde{t}_2},\theta_{\tilde{t}}) \sim 0$), both inclusive $pp \to h$ and $pp \to h + \text{jet}$ processes will appear SM-like.%, up to corrections of the order $v^2/m_t^2$. and the effect of $F_g\sim m_t^/m_{\tilde{t}}^2$ \adam{still confused by this $v/m_t$ is not small...I also dont understand the 'effect of $F_g$'}.

At intermediate $p_T, m_t < p_T < m_{\tilde t_i}$ we can approximate the amplitude as the sum of a high-$p_T$ piece (Eq.~\eqref{eq:hipt}) for the top plus a decoupled piece (Eq.~\eqref{eq:Mppp-stop-lowpt}) for the top squarks:
\begin{equation}
  \label{eq:M+++-intpt}
  \left.\mathcal{M}_{+++}\right|_{m_t \ll p_T\ll m_{\tilde t_1},m_{\tilde t_2}} \frac{m_t^2}{p_T}
  \left(A_0+A_1\ln\left(\frac{p_T^2}{m_t^2}\right)+A_2\ln^2\left(\frac{p_T^2}{m_t^2}\right)\right)+ 8 \frac{\Delta}{p_T^2}\,F_g\,.
\end{equation}
The contribution of the top quarks is again proportional to $F_g$ -- so if
there is a cancellation between two top squark contributions in the total cross
section it will persist in this regime.

 To break the cancellation, we must go to $p_T$ higher than the mass of the lighter stop. Here, we can approximate the amplitude as a high-$p_T$ contribution from the top loop and lightest top squark loop, and a decoupled piece for the heavier top squark:
\begin{multline}
  \label{eq:M+++-highpt}
    \left.\mathcal{M}_{+++}\right|_{m_t,m_{\tilde t_1} \ll p_T\ll
      m_{\tilde t_2}}\simeq \frac{m_t^2}{p_T}
    \left(A_0+A_1\ln\left(\frac{p_T^2}{m_t^2}\right)+A_2\ln^2\left(\frac{p_T^2}{m_t^2}\right)\right)\\
    + \frac{g_{h\, \tilde{t}_i \tilde{t}_i}\, v}{p_T}\left(\tilde
      A_0+\tilde A_1\ln\left(\frac{p_T^2}{m^2_{\tilde t_1}}\right)+\tilde
      A_2\ln^2\left(\frac{p_T^2}{m^2_{\tilde t_1}}\right)\right)-\frac{4}{3}\frac{\Delta}{p_T^2}
    \frac{g_{h\, \tilde{t}_i \tilde{t}_i}\, v}{m_{\tilde t_2}^2} \,.
\end{multline}
The couplings $g_{h\tilde t_i\tilde t_i}$ multiply different kinematic functions rather than combining into $F_g$, so the amplitude is sensitive to the top squarks even when parameters conspire to make $F_g \sim 0$. Note that any cancellation between the contributions of the stops is
also broken at very high $p_T$, i.e.~$p_T\gg m_{\tilde t_1}, m_{\tilde t_2}$. 

To summarise,  we have shown that by looking at the high-$p_T$ behavior of $h+\text j$ once can break model degeneracies in the top squark sector,  opening up sensitivity to parameter space that more inclusive searches cannot probe. However, the added information in $h + \text{j}$ comes only when we consider $p_T$ higher than the mass of the lightest top squark. %\adam{transition is awk}
Whether or not one can expect to reach such kinematic $p_T$ regime at the LHC depends on the physical masses $m_{\tilde t_{1,2}}$. To get a more quantitative idea of the size of the deviation top squarks can cause, we turn to numerics.

%o get a more quantitative idea of  how much where we perform a numerical analysis of these contributions and compare with the reach from inclusive Higgs production. 

%After gaining analytical understanding of the different $p_T$ regimes in the $h$+jet spectrum, and the regime where model degeneracies could be broken, we move onto the next section where we perform a numerical analysis of these contributions and compare with the reach from inclusive Higgs production. 

%%%%%%%%%%%%%%%%%%%%%%%%%%%%%% 
%%%%%%%%%%%%%%%%%%%%%%%%%%%%%% 
\section{Numerical Results}
%%%%%%%%%%%%%%%%%%%%%%%%%%%%%% 
%%%%%%%%%%%%%%%%%%%%%%%%%%%%%% 
\label{sec:mcresult}

Using the squared matrix elements in Eqs.~\eqref{eq:squaredamp}
and~\eqref{eq:squaredampquarks}, we compute the $p_T$ spectrum of the
Higgs $d\sigma/dp_T$. The actual calculation of the $p_T$ spectra
results from interfacing a modified version of {\tt
  HERWIG}~\cite{herwig} with the parton density toolkit {\tt
  HOPPET}~\cite{hoppet}. Our results have been validated against the
existing program {\tt SusHi}~\cite{sushi}, using the MSSM input card,
with very large masses for the bottom squarks and the gluinos. The main
difference between our implementation and {\tt SusHi} is that we
compute the full $p_T$ spectrum with a Gaussian integrator in a single
run, whereas {\tt SusHi} uses a Monte-Carlo integrator to provide a
single $p_T$-bin for each run. In terms of performance, with a single current CPU, with our implementation one can obtain the entire $p_T$-spectrum for a mass point in less than a second, whereas to run {\tt SusHi} in one single $p_T$-bin would take about a minute.
%comparison \adam{The way its written, it looks like everything we did is already in SusHi. Can we elaborate more on what we did thats new?}

\begin{figure}[htbp]
      \centering
      \includegraphics[width=.8\textwidth]{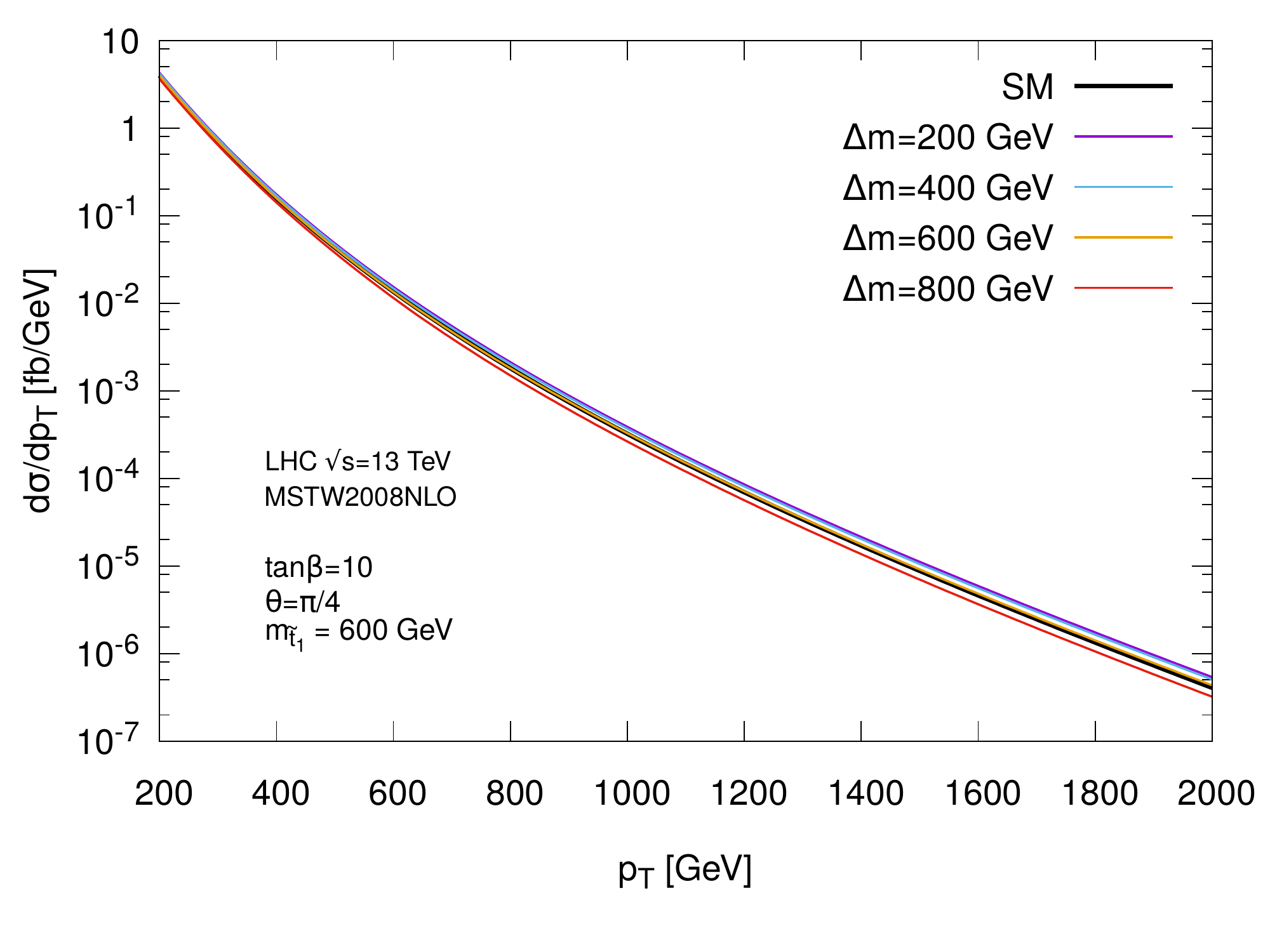}
      \caption{Transverse momentum spectra of the Higgs boson in the
        Standard Model, and for four different values of the mass
    difference $\Delta m$. See text for details.}
  \label{fig:dsdpt}
    \end{figure}

In Fig.~\ref{fig:dsdpt}, we show numerical results
for $d\sigma/dp_T$ for $m_{\tilde t_1}=600\,$GeV and four different
values of the mass difference $\Delta m$, obtained with the
MSTW2008NLO parton distribution set~\cite{Martin:2009iq}, for $\tan\beta=10$ and
maximal mixing, i.e.\ $\theta=\pi/4$, as well as the corresponding
prediction in the SM. All distributions have been
obtained by setting both renormalisation scale $\mu_R$ and
factorisation scale $\mu_F$ equal to $(p_T+\sqrt{p_T^2+m_H^2})/2$. We
note immediately that, for the chosen parameters, the difference
between the SM spectrum and that with additional top squarks in the loops is
not huge, at most 30\% in the highest $p_T$ bins. The smallness of the effect is expected from the analytical results in Sec.~\ref{sec:H1j}, Eqs.~(\ref{eq:Mppp-stop-lowpt}-\ref{eq:M+++-intpt}). 

We can compare these results with the contributions of fermionic top-partners to the same process discussed in Ref.~\cite{Banfi:2013yoa}, in particular the high-$p_T$ behavior described in Eq.~(4.3) of that paper, where the dependence on $m$, the scale of new physics, goes  as $m^2/p_T$, instead of $m_t^2/p_T$. Therefore, one would typically expect more sizeable effects from new fermionic top-partners than from stops.  
%\adam{Would be great to say something here about the difference with the top partner case.  }

In order to better assess deviations from the SM behaviour in the spectra, we construct the cumulative distribution $\sigma(p_T^{\rm cut})$, defined by
\begin{equation}
\label{eq:sigmacut}
\sigma(p_T^{\rm cut}) = \int_{p_T^{\rm cut}}^\infty \! dp_T\,\frac{d\sigma}{dp_T}\,,
\end{equation} 
and we consider the deviation $\delta(p_T^{\rm cut})$ of the cumulative
cross section $\sigma(p_T^{\rm cut})$ from its expected value in the
Standard Model as follows:
\begin{equation}
\label{eq:delta}
  \delta(p_T^{\rm cut})=\frac{\sigma(p_T^{\rm cut})-\sigma^{\rm SM}(p_T^{\rm cut})}{\sigma^{\rm SM}(p_T^{\rm cut})}\,.
\end{equation}
\begin{figure}[htbp]
  \centering
  \includegraphics[width=.65\textwidth]{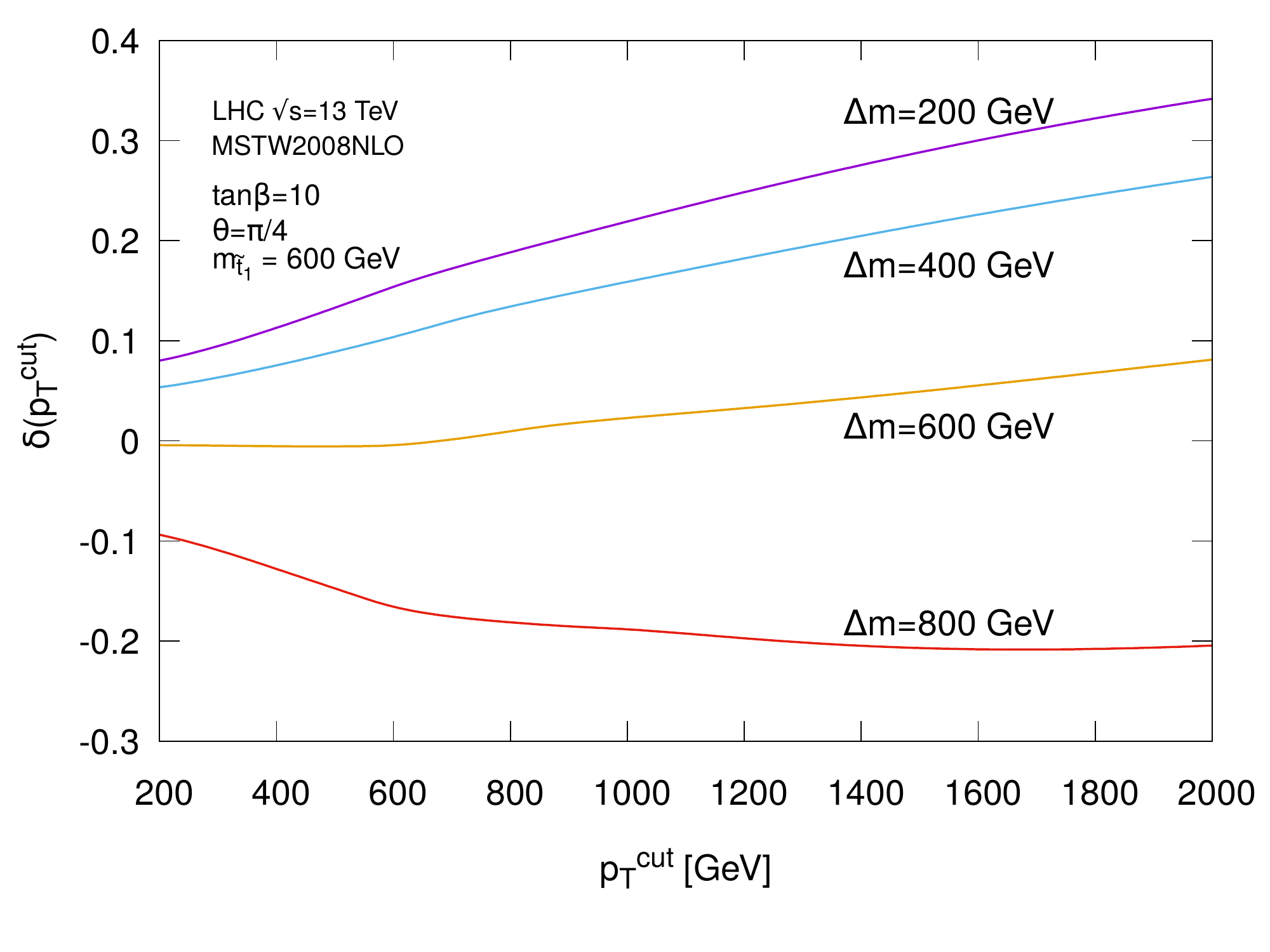}
  \caption{The deviation from SM expectation defined in
    Eq.~\eqref{eq:delta} as a function of $p_T^{\rm cut}$ for
    $m_{\tilde t_1}=600\,$GeV and four different values of the mass
    difference $\Delta m$.  See the main text for details.}
  \label{fig:delta}
\end{figure}
\begin{figure}[htbp]
  \centering
    \includegraphics[width=.65\textwidth]{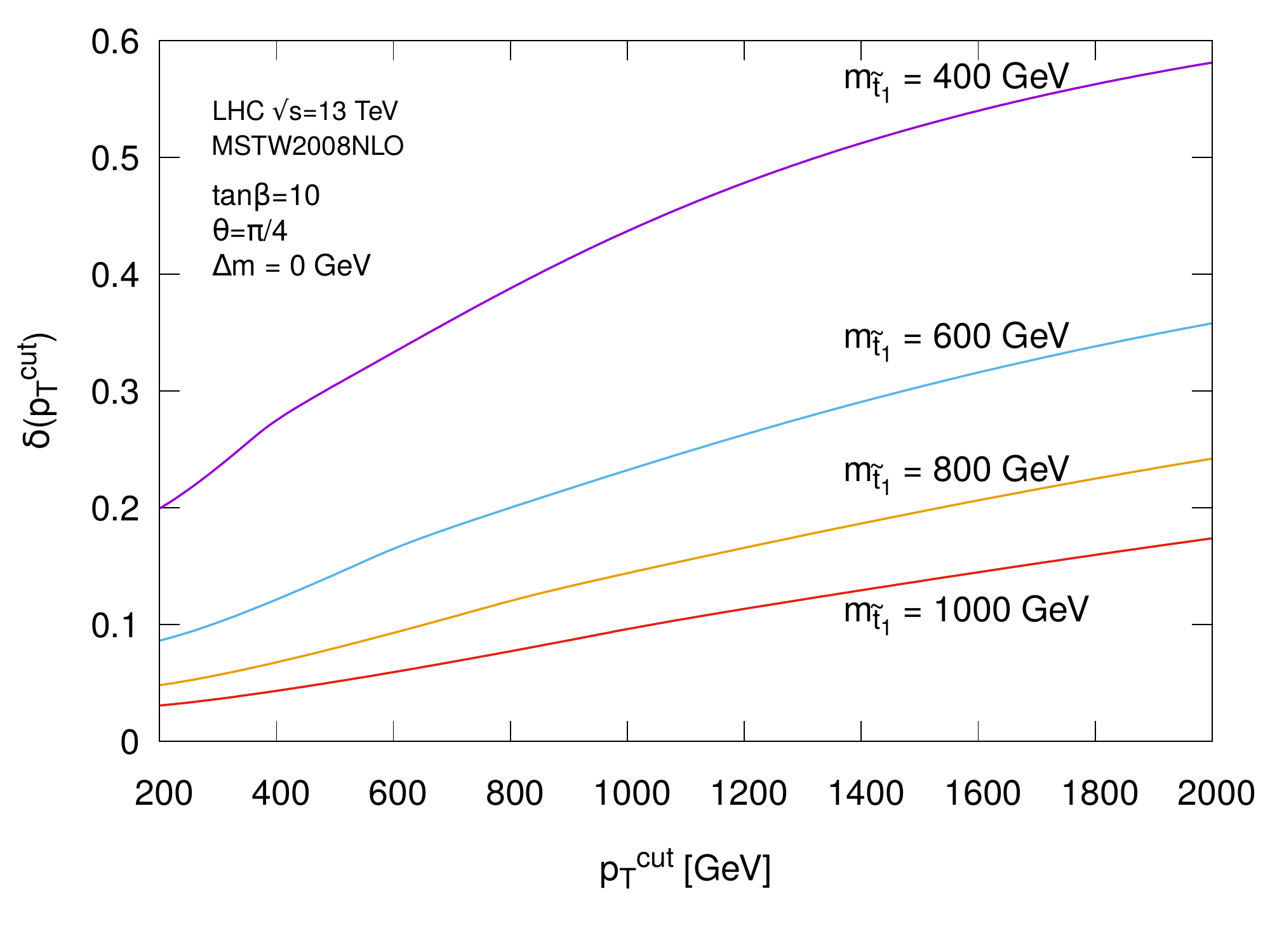}
  \caption{The deviation from SM expectation defined in
    Eq.~\eqref{eq:delta} as a function of $p_T^{\rm cut}$ for
   $\Delta m$=0 (degenerate stops) and for different values of the stop masses.  See the main text for details.}
  \label{fig:delta2}
\end{figure}
The $\delta$ values corresponding to the parameters used in
Fig.~\ref{fig:dsdpt} are shown in Fig.~\ref{fig:delta}. Note that each
prediction consists only of a single curve obtained by fixing
$\mu_R=\mu_F=(p_T+\sqrt{p_T^2+m_H^2})/2$ in each $p_T$ spectrum. We
have checked that a simultaneous variation $\mu_R$ and $\mu_F$ in the
range $1/2\le \mu_R/\mu_F \le 2$ in the numerator and denominator in
Eq.~\eqref{eq:delta} has a negligible impact on
$\delta(\ptcut)$. Therefore, no band associated with such scale
variations is shown in Fig~\ref{fig:delta}.

Let us now comment on the robustness of the variable $\delta$ respect
to the introduction of NLO effects. Recently, a complete calculation
of the differential Higgs+jet process at NLO with fulll top-mass
effects has been performed~\cite{newNLO},
see also~\cite{Lindert:2018iug,Neumann:2018bsx} for recent predictions with
an approximate treatment of the top-quark mass dependence. The
authors in Ref.~\cite{newNLO} find that the K-factor (the ratio
NLO/LO) for this process is roughly 2 and quite flat with
$p_T$. Hence, one could simply re-scale the SM contribution in
Eq.~\eqref{eq:delta} by this K-factor, improving the SM prediction to
NLO. But the equivalent calculation for the BSM contribution is not
known. Nevertheless, one can expect that the K-factor of the BSM
contribution will be of similar size as the SM one. In fact, on one
hand, one can argue that for large stop masses, the BSM contributions
behaves in a similar way as the heavy-top effective theory
($m_t\to \infty$), which has a K-factor of about 2. On the other hand,
Ref.~\cite{newNLO} also shows that for finite masses the K-factor one
obtains is of the same order as that of the heavy-top effective
theory. Therefore, one would then expect a similar K-factor (flat with
$p_T$ and of order 2) in the stop contribution to $\sigma$, and hence
our calculation of the quantity $\delta$ and LO should be similar to
the NLO.  This suggests that this variable may be robust against
higher-order contributions, although this statement will have to be
validated when NLO corrections to the BSM $p_T$ spectrum will become
available. In the next section we show plots with $\delta$ at LO, and
the reader should keep in mind the caveats mentioned before.

%\adam{Where is this? isnt this in the next section?? }
%\adam{redo, with analytics done earlier,
%  replace text here with references back to those equations to explain
%  low, high pT behavior}

\section{ Comparing inclusive Higgs and  high-$p_T$ Higgs sensitivities }
\label{sec:comparing_sensitivity}

Having reviewed the top squark contributions to inclusive Higgs production and Higgs plus jet, we now compare the LHC sensitivity in the two modes. For our comparison, we will fix $\theta$ and $\tan{\beta}$, and plot our results as contours in the $m_{\tilde t_1}, \Delta m$ plane. For $\theta$, we pick two benchmarks, $\theta = 0, \theta = \pi/4$; these correspond to the extremes of no mixing and maximal mixing among the different top squarks -- other choices of mixing angle would fall between the two. 

 We now discuss the range of parameters that can be probed through the
variable $\delta$, by showing different contours for
$\delta(p_T^{\rm cut})$ (translated into a percent deviation) as a
function of the mass of the lightest top squark $m_{\tilde t_1}$ and
the top squark mass difference $\Delta m$. The filled contours
correspond to $\delta(p_T^{\rm cut})$, as defined in
Eq.~\eqref{eq:delta}, whereas the dashed lines represent analogous
contours for the total cross section, obtained from
eq.~\eqref{eq:rgludef}. All the results we show
correspond to $\tan\beta=10$. We have checked that other values of
$\tan\beta$ lead to similar results. Furthermore, all the parameters
we have considered are not excluded by present data of the Higgs $p_T$
spectrum~\cite{Aad:2015lha}.  

\begin{figure}[htbp]
\includegraphics[width=0.5\textwidth]{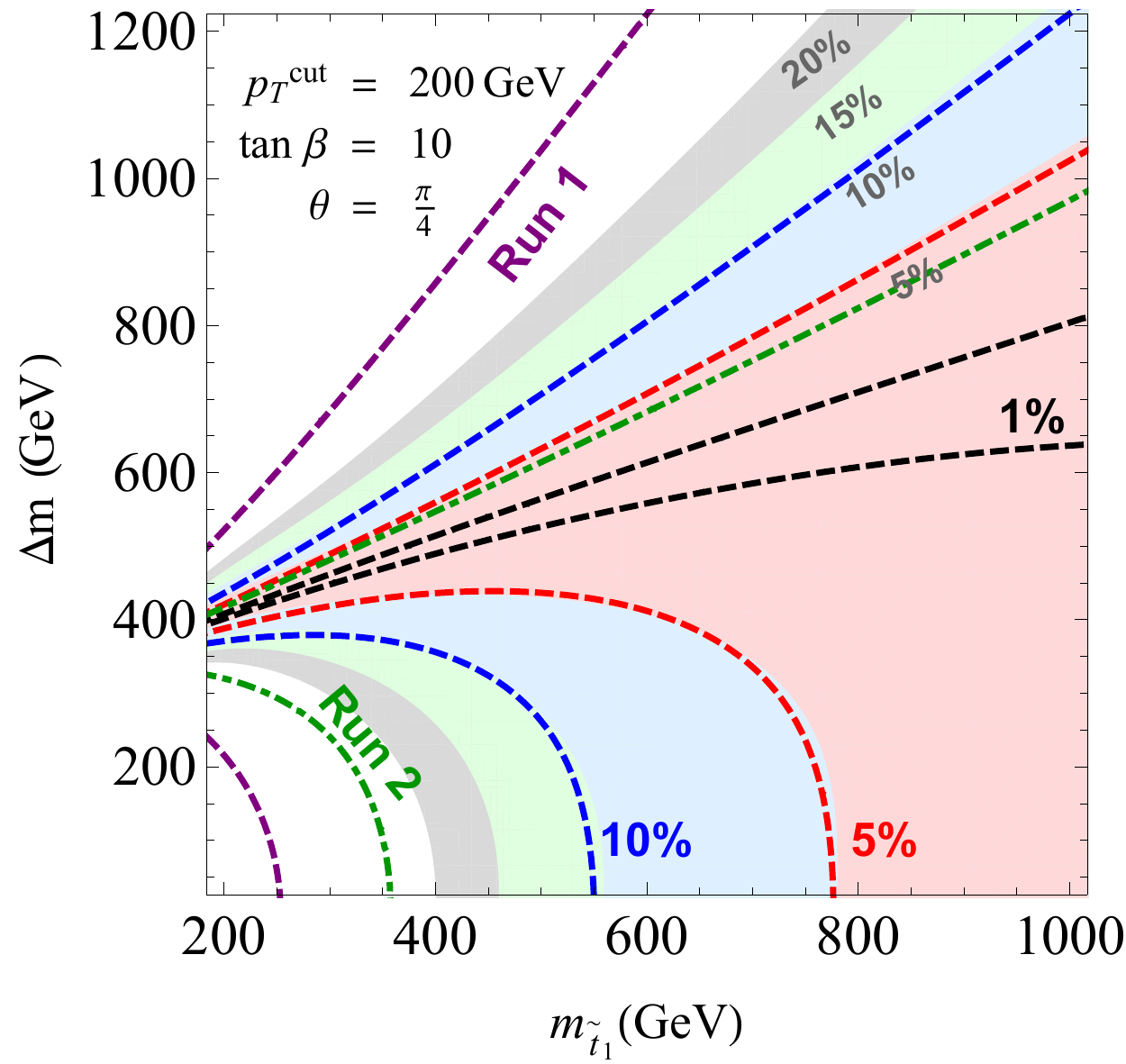}
\includegraphics[width=0.5\textwidth]{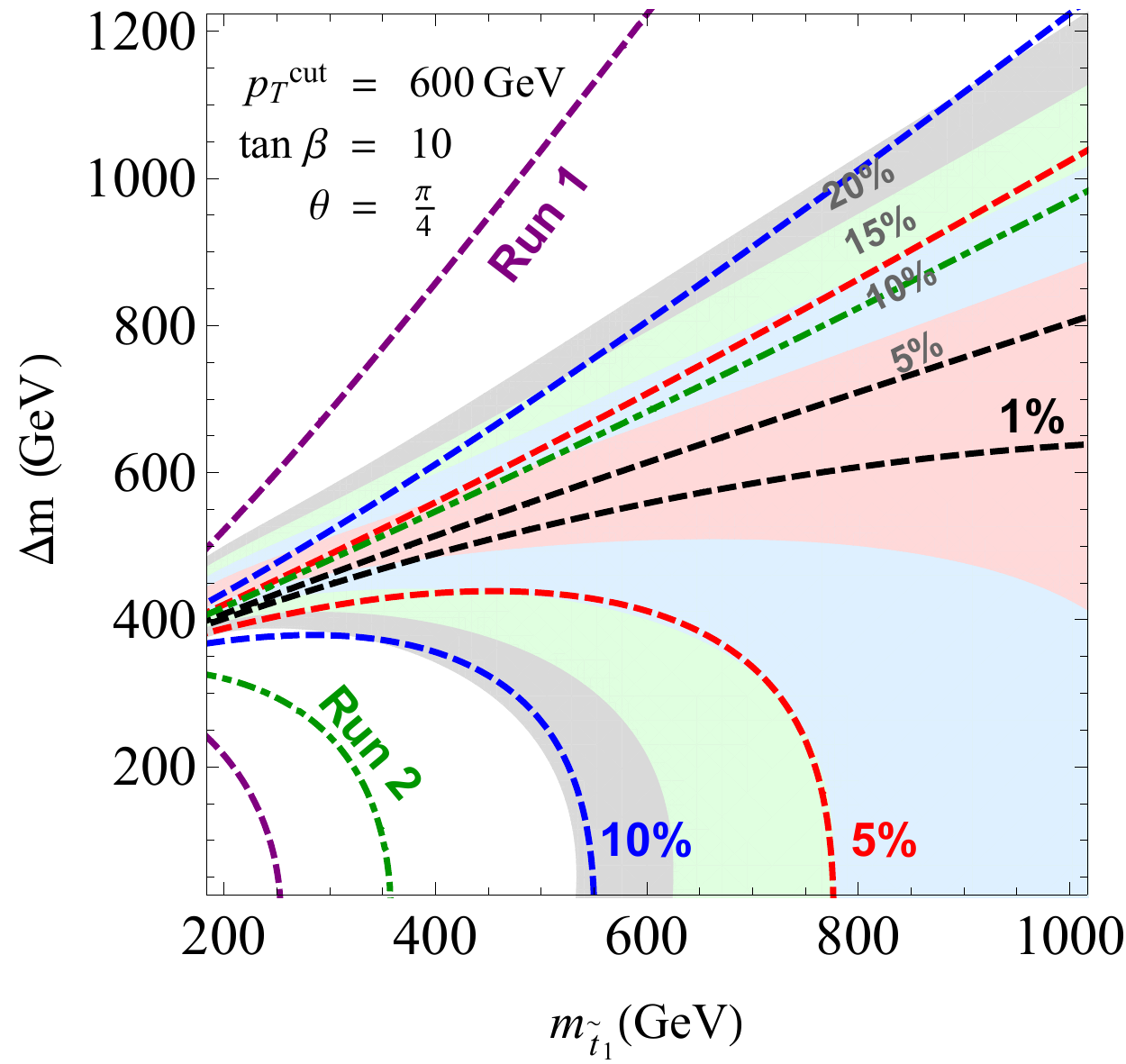}
\caption{Contour plots for $\delta$ (as percent deviation) for the
  integrated Higgs $p_T$ spectrum (solid) and for the Higgs total cross
  section (dashed), as a function of the lighter top squark mass $m_{\tilde
    t_1}$ and of the top squark mass difference $\Delta m$, for $\tan\beta=10$,
  $\theta=\pi/4$, and two different values of $p_T^{\rm cut}$, namely
  $p_T^{\rm cut}=200\,$GeV and $p_T^{\rm cut}=600\,$GeV.}
\label{fig:delta-sensitivity-thpi4}
\end{figure}

\begin{figure}[htbp]
\includegraphics[width=0.5\textwidth]{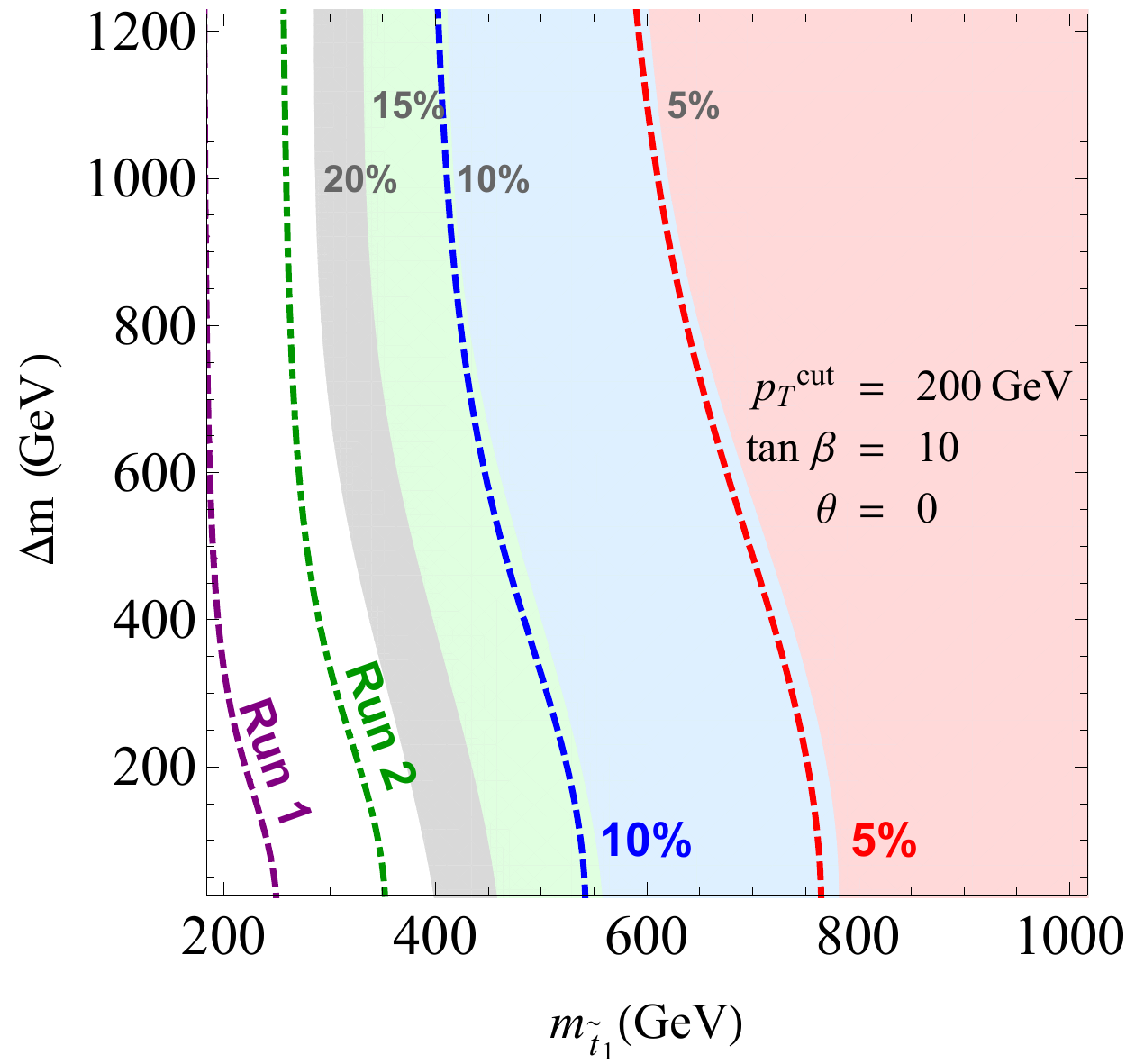}
\includegraphics[width=0.5\textwidth]{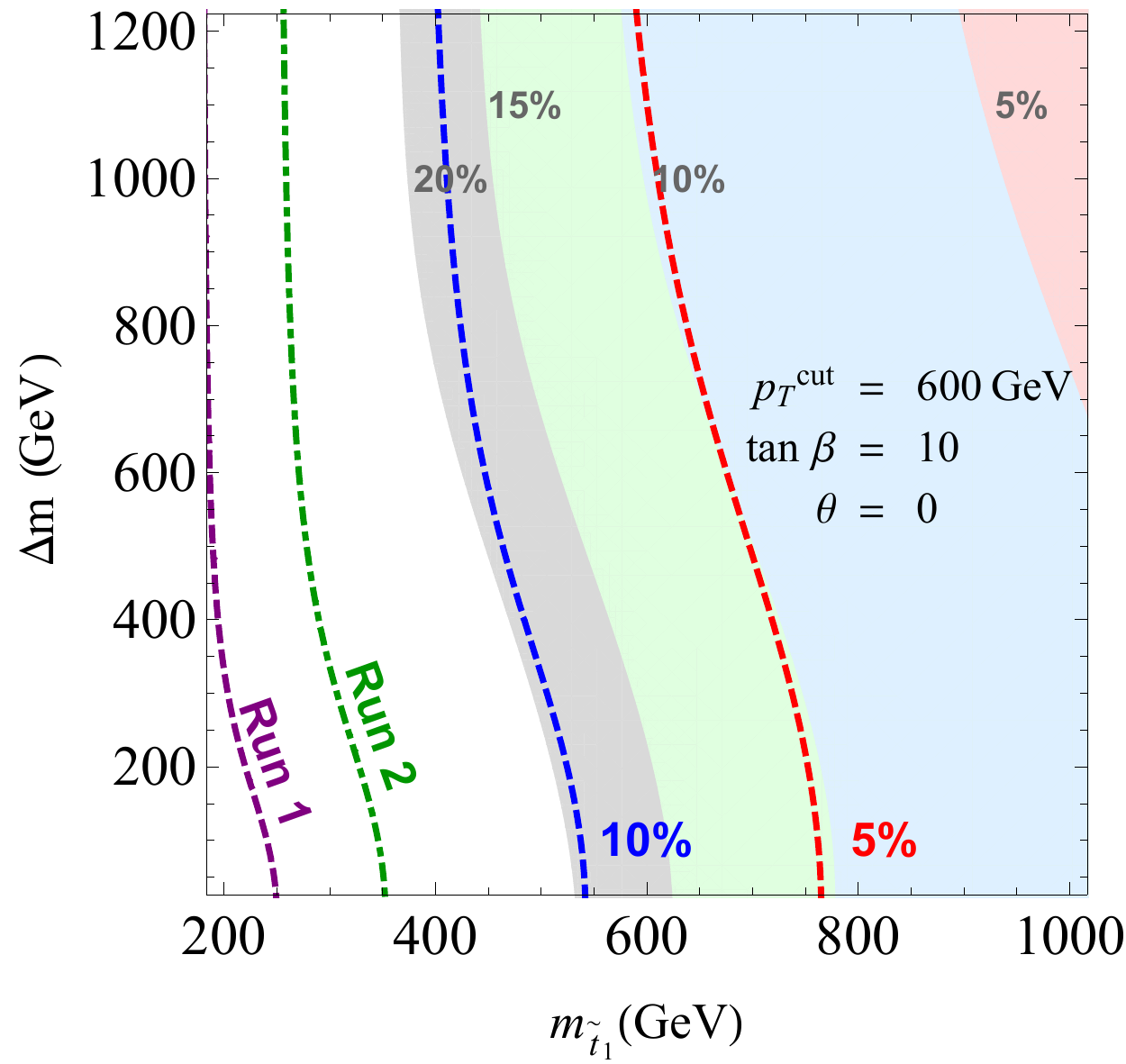}
\caption{The same contour plots as in
  Fig.~\ref{fig:delta-sensitivity-thpi4}, but for $\theta=0$.}
\label{fig:delta-sensitivity-thpi0}
\end{figure}

For inclusive Higgs production, each value of $\theta, m_{\tilde t_1}, \Delta m$ can be mapped to $\kappa_g$ following Eq.~\eqref{eq:rgludef} and \eqref{eq:Fgdef}. Via $\kappa_g$, each parameter point maps onto an inclusive Higgs cross section which can then be compared to LHC limits (both current and projected). The result is shown in dashed lines in Figs.~\ref{fig:delta-sensitivity-thpi4} and \ref{fig:delta-sensitivity-thpi0}. Regions to the left of the purple dashed line (green dashed line) are excluded by LHC Run-I \cite{TheATLASandCMSCollaborations:2015bln} (LHC Run-II \cite{ATLAS:2017ovn}) data.  The blue, red, and black dashed lines show future sensitivity, quoted in terms of the percent deviation in the inclusive cross section coming from the top squarks (for the future bounds, the region to the left is excluded). As the difference between the two frames in Figs.~\ref{fig:delta-sensitivity-thpi4} and \ref{fig:delta-sensitivity-thpi0} applies only to the Higgs plus jet mode, the dashed lines are the same in both panels. When the two top squarks are highly mixed we can easily spot the parameter region where a cancellation occurs between them: for $\theta = \pi/4$, the cancellation occurs along a line between $(m_{\tilde t_1}, \Delta m) = (200\, \gev, 400\,\gev)$ and extending to $(1000\, \gev, 700\,\gev)$. When $\theta = 0$, there can be no cancellation between top squarks and the bounds in $(m_{\tilde t_1}, \Delta m)$ space look qualitatively different.

The inclusive Higgs production contours are overlayed on top of Higgs plus jet $\delta (p_T^{\rm cut})$ contours, with $\delta(p_T^{\rm cut})$ defined in Eq.~\eqref{eq:delta}. In the left panel, we show $\delta(200\, \gev)$ while the right panel we show $\delta(600\, \gev)$; in both frames we show $\delta$ deviations of $5, 10, 15$ and $20\%$. Focusing first on the maximal mixing case (Fig.~\ref{fig:delta-sensitivity-thpi4}) and comparing the two panels, we can see the impact of the Higgs $p_T$ cut. For $\ptcut = 200\, \gev$, we have $p_T^{\rm cut}\lesssim m_{\tilde t_1}$ so the  the top squarks can
be considered as decoupled in the bins of the Higgs $p_T$ spectrum
that contribute most to $\delta(p_T^{\rm cut})$. As
shown in Eq.~\eqref{eq:M+++-intpt}, the top squark contribution in this regime is proportional to the same term $F_g$ appearing in
eq.~\eqref{eq:Fg}. Therefore, whenever there is a cancellation in the total
Higgs cross section, there will also be a cancellation in $\delta(p_T^{\rm cut})$. This cancellation can also be appreciated by looking at the
curves in Fig.~\ref{fig:delta}. Picking to $\Delta m=600\,$GeV as an example top squark mass, we see that we need to increase $p_T^{\rm
  cut}$ to more than $600\,$GeV to see an appreciable deviation of
$\delta(p_T^{\rm cut})$ from one.  In fact, for $p_T^{\rm
  cut}>600\,$GeV we start to open at least the loop containing the
lighter top squark. This slightly larger sensitivity then
is reflected in the right panel of
Fig.~\ref{fig:delta-sensitivity-thpi4}, as the e.g. $\delta = 10\%$ contour cuts out more parameter space than the $10\%$ contour for inclusive Higgs production (dashed blue) for $m_{\tilde t_1} \lesssim 400\, \gev$.

 In Fig.~\ref{fig:delta-sensitivity-thpi0}, we show the analogous plots for the case in which no mixing occurs,
i.e.\ $\theta=0$. As in the maximal mixing scenario, when $\ptcut=200\,$GeV both the
Higgs cross section and the Higgs $p_T$ spectrum have very similar
sensitivities. This picture changes when increasing $\ptcut$ to
$600\,$GeV, where the sensitivity of the Higgs $p_T$ spectrum is
essentially doubled with respect to that of the total cross
section. This increase in mass sensitivity as one increases the cut is expected from the analytical analysis and can be seen in Fig.~\ref{fig:delta2}.  

As there is no possible cancellation among top squarks when $\theta = 0$, $F_g < 0$ and the amplitudes for $pp \to h$ and $pp \to h + \text{jet}$ are always increased by new physics. 
In this way, the $\theta = 0$ case is the scalar analogue of the
contribution of a fermionic partner of the top presented
in~\cite{Banfi:2013yoa}. However, the sensitivity of the $p_T$ spectrum to contributions from
fermionic top partners found in~\cite{Banfi:2013yoa} is much larger than the sensitivity to scalars we find here. This difference is due to
the specific interplay between a top and a top-partner in composite
Higgs models for different values of the Higgs transverse momentum. In the case of composite Higgs models, there is a cancellation between the top contribution and the top partner contribution which occurs
whenever both states are decoupled. As the $p_T$ of the Higgs increases above $m_t$, a
heavy top-partner stays decoupled, while the top quark behaves as a light
particle, breaking the cancellation between contributions. This is
not the case for top squarks, where there is a region of parameter space where the contribution of the two top squarks cancels when they are both
decoupled, while the top quark contribution remains SM-like. Therefore,
in order to break the cancellation, one needs to reach transverse
momenta that exceed the mass of the lighter top squark. Furthermore, due to
the fact that a top partner is a chiral fermion while a top squark is a
scalar, the contribution to $pp \to h + \text{jet}$ from a heavy top squark is suppressed by one extra
power of the top squark mass. 

Finally, note that in this paper we are using Higgs data to indirectly probe top squarks. Currently, indirect top squark bounds are in the range of $m_{\tilde t}\sim$ 300 GeV~\cite{nsusy, Johnlatest}. The most recent direct searches for stops reach the TeV region, however this is not a fair comparison as direct searches are based on the assumption that stops decay into specific final states, largely involving missing energy signatures~\cite{stopsearch}.

\section{Discussion and Outlook}
\label{sec:conclude}

In this paper we studied how new colored scalars, top squarks, could affect the production of the Higgs boson.  We were particularly interested in the interplay between inclusive gluon-fusion and differential Higgs production, $h$+jet. 

At first glance, one would think the inclusive production $g \, g \to h$  should be the dominant handle on possible new coloured particles. This production enjoys rich statistics and a solid experimental and theoretical understanding. The effect of new physics, however, could be reduced due to symmetries (such as in many models in Composite Higgs) or simply due to accidental cancellations (as in classes of SUSY sectors). Under these circumstances, the study of differential rates of the Higgs production in association with a high-$p_T$ jet is then the best handle to uncover new physics. 

In prior work, Ref.~\cite{Banfi:2013yoa}, we studied the sensitivity of the $h$+jet channel in the quite dramatic case of fermionic top-partners in composite Higgs models, where low-energy theorems tend to protect the inclusive production from any variations from the SM; see also Refs.~\cite{Azatov:2013hya, Azatov:2013xha}.  

In this paper we have focused on scalar top-partners, and in particular on supersymmetric top squarks, where no such low-energy theorems are present. Nevertheless, we found that there is an interesting interplay between the information contained in the $h$+jet differential production and the inclusive production.  We presented analytical expressions for the corresponding amplitudes including the dependence on the stop spectrum,  as opposed to previous studies~\cite{Hollik,sushi,Grojean:2013nya}.

Obviously, if an accidental cancellation occurs in the stop sector leading to a reduced gluon fusion rate~\cite{Grojean:2013nya}, the differential production could become the best handle to discover new physics.  Even if no dramatic cancellation occurs, the $h$+jet rate still adds value to the search for new physics. Searches in gluon-fusion inclusive and the boosted Higgs topologies face very different background challenges. Indeed, as we have seen in searches by ATLAS and CMS, some decays of the Higgs may be more accessible in the boosted regime than in the inclusive case. Hence, a combined analysis of the two topologies would provide us with the best handle to {\it dig} top squarks from the LHC data. 

We have obtained analytical expressions for the  $h$+jet amplitude in various $p_T$ regimes and found that the information one could gain on stops using differential rates is, unsurprisingly, concentrated in the regime $p_T \gtrsim m_{\tilde t}$. We then performed a numerical study to evaluate these relative effects. As a simple measure of the differential rate, we have defined a cumulative variable, namely the excess of events above a certain bin in $p_T$ with respect to the SM, $\delta(p_T)$. We argued that $\delta$ may be more robust against theoretical and experimental uncertainties than a fully differential study. With this naive measure of new physics in the high-$p_T$ region, we chose two benchmark values $p_T>$ 200 and 600 GeV and compared future prospects for inclusive and differential information under some assumptions regarding the level of uncertainties for each topology.  As expected, larger cuts on $p_T$ can lead to increased sensitivity, but the gain has to be weighted against the loss of statistics. This is a similar situation encountered when using the missing-energy distribution in searches for SUSY Dark Matter~\cite{Barducci:2015ffa} and a similar detailed analysis should be done.  Such an analysis in the case of Higgs+jet topologies is feasible (at LO) with current Monte-Carlo event generators, through \texttt{aMCSusHi}, the interface of the fixed-order program SusHi  to aMC@NLO~\cite{Mantler:2015vba}.  

In order to completely assess which values of $\delta$ can be actually
probed by experiments we need to be able to accurately determine the
SM contribution. This requires considering all possible backgrounds to
Higgs production in the selected decay channels, e.g. $h\to \gamma\gamma$ or to four-leptons, estimating the
associated systematic uncertainties and performing a suitable
statistical data analysis. Such an study is beyond the scope of this paper, but one can examine the results in Ref.~\cite{Schlaffer:2014osa}, where the authors considered the effect of $\kappa_g$ as an effective operator (see also Ref.~\cite{Harlander:2002wh} for an NLO analysis). Specifically, using the transverse momentum spectrum of a Higgs decaying into $\tau\tau$ and $W W^*$,  recoiling against a
jet with $p_{t,j}>200\,$GeV, and $3\,\mathrm{ab}^{-1}$ of luminosity
at the HL-LHC, Ref.~\cite{Schlaffer:2014osa} claims that it is
possible to exclude at 95\% confidence level values of $\kappa_g$ in
the range $\kappa_g<-0.4$ and $\kappa_g>0.3$, with the additional
constraint that no deviation is seen in the Higgs total cross section,
and assuming that experimental systematic uncertainties are at most
10\% \footnote{The prospects of measuring the coupling of the Higgs to gluons in future lepton colliders are promising. For example, $\Delta BR (h\to g g )/BR(h\to g g)$ = 4\% and 7\% for ILC250, ILC500~\cite{ILCTDR}.}. For $p_{t,j}>200\,$GeV, these values of $\kappa_g$ correspond to
roughly a 6-7\% deviation from the SM. This means that obtaining a
similar sensitivity in the present case is not unreasonable. 

However, the analysis in Ref.~\cite{Schlaffer:2014osa} relies on the dramatic growth with energy of the higher-dimensional operator, which in turns results in deviations from the SM on the Higgs transverse momentum spectrum which become as big as 80\%  for $p_T>600\,$GeV.  This is quite different from our situation,
where increasing the cut on the jet transverse momentum does not lead
to huge deviations from the SM. Indeed, let us compare the prospects drawn in Ref.~\cite{Schlaffer:2014osa} with a simple  case where one single stop dominates the phenomenology. Namely, let us take Eqs.~\eqref{kappaap} and \eqref{expF} in the case $m_{\tilde t_1} \ll m_{\tilde t_2}$ and $\theta$=0, leading to $\kappa_g \simeq 1+ \frac{4 m_t^2}{m_{\tilde t_1}^2}$. A 10\% reach in $\kappa_g$ would then mean a limit $m_{\tilde t_1}>$ 1 TeV,  which is clearly beyond the sensitivity we expect when looking at the full stop contributions. 

In conclusion, we hope this paper serves to motivate the experimental collaborations to perform a combined analysis of gluon-fusion and differential information to search for new physics. We have provided an analytical understanding of the differential rates in various regimes of $p_T$ and defined a useful, but rather simplistic, variable $\delta$ to encompass some of the differential information. We also hope to  encourage theorists to perform the calculation of the differential distribution for stops at NLO QCD, which would be ultimately needed to sensibly compare with the SM predictions in the high-$p_T$ tails. 

\section*{Acknowledgements}

The work of AB and VS is supported by the Science Technology and Facilities Council (STFC) under grant number ST/P000819/1. The work of AM was partially supported by the National Science Foundation under Grant No. PHY-1520966. 

\appendix

\section{Higgs plus one jet for large top squark masses.}
\label{sec:H1j-decoupling}

Here we give analytical expressions for the helicity amplitudes
introduced in section~\ref{sec:H1j} in the ``decoupling'' limit $m^2\gg
m_H^2,s,|t|,|u|$, where $m$ is the scalar running in the loops. 

First, we give the expansion of the scalar integrals appearing in the
amplitudes:
\begin{equation}
  \label{eq:scalar-decoupling}
  \begin{split}
  & B_1(q^2) \simeq \frac{q^2-m_H^2}{6 m^2}\,,\quad
  C_1(q^2) \simeq -\frac{1}{2 m^2}-\frac{q^2+m_H^2}{24 m^4}\,,\\
  & D_0(s,t)  \simeq \frac{1}{6m^4}\,,\qquad E_0(s,t)\simeq\frac{u}{m^2}\,.
  \end{split}
\end{equation}
This gives
\begin{equation}
\label{eq:M-decouple}
\begin{split}
  M^{\tilde t_i}_{+++} &\simeq -\frac{\Delta g_{h\, \tilde{t}_i \tilde{t}_i}
    v}{stu}\frac{4}{3}\frac{s^2}{m^2}\,,\qquad 
 M^{\tilde t_i}_{++-}
  \simeq \frac{\Delta g_{h\, \tilde{t}_i \tilde{t}_i} v}{stu}\frac{4}{3}\frac{m_H^4}{m^2}\,,\\
  M^{\tilde t_i}_{-+-} & \simeq -\frac{\Delta g_{h\, \tilde{t}_i \tilde{t}_i}
    v}{stu}\frac{4}{3}\frac{t^2}{m^2}\,,\qquad 
  M^{\tilde t_i}_{-++} \simeq -\frac{\Delta g_{h\, \tilde{t}_i \tilde{t}_i}
    v}{stu}\frac{4}{3}\frac{u^2}{m^2}\,.  
\end{split}
\end{equation}
Similarly,
\begin{equation}
  \label{eq:Mq-decoupling}
  \mathcal{M}^{\tilde t_i}(q\bar q\to hg)\simeq -\frac{g_{h\, \tilde{t}_i \tilde{t}_i} v}{24}\frac{s_1}{m^2},
\quad  \mathcal{M}^{\tilde t_i}(qg\to hq)\simeq -\frac{g_{h\, \tilde{t}_i \tilde{t}_i} v}{24}\frac{u_1}{m^2} \, ,
\end{equation}
\begin{equation}
 \mathcal{M}^{\tilde t_i}(gq\to hq)\simeq -\frac{g_{h\, \tilde{t}_i \tilde{t}_i} v}{24}\frac{t_1}{m^2}\,.
\end{equation}

\section{Higgs plus one jet in the soft and collinear limit. }
\label{sec:H1j-soft-coll}

In this appendix we report the soft and collinear limits of the
amplitudes and matrix elements for Higgs plus one-jet production
computed in section~\ref{sec:H1j}. We believe this information might
be useful for future studies aiming at exploiting the analytical
properties of the matrix elements.

The main property of soft and collinear limits of matrix elements is
that they factorize into the product of the tree-level matrix element
and universal functions. Therefore, we first need the expression of
the Born matrix element. Due to conservation of angular momentum, the
amplitude for the process $gg\to h$ is non-zero only if the two gluons
have the same helicity, say both positive.
 The un-averaged matrix element squared for this process is
\begin{align}
| M_{gg\to h}|^2 &= \frac{(N_c^2 - 1) \alpha_s^2}{32 \pi^2 v^2} \ \left| \sum_{i=t,b,\tilde{t}_1,\tilde{t}_2} \mathcal M_{++}^{i} \right|^2 \,.
\label{eq:squaredborn}
\end{align}
The top squark contribution to the above equation is
\begin{equation}
  \label{eq:Born-amplitude}
  \mathcal{M}^{\tilde{t}_i}_{++} = 2 g_{h\, \tilde{t}_i \tilde{t}_i} v(1+ 2 m^2 C_0(m_H^2))\,,
\end{equation}
with $g_{h\, \tilde{t}_i \tilde{t}_i}$ is either of the couplings defined in equations~\eqref{eq:ghtt}, \eqref{eq:ghtt2}.

\subsection{Soft limit}
\label{sec:soft}

The soft limit $p_3 \to 0$ corresponds to 
\begin{equation}
  \label{eq:soft-limit}
  s \to m_H^2\,,\quad u,t\to 0\,,\quad s_1\to 0\,,\quad u_1,t_1\to -m_H^2\,.
\end{equation}
Keeping the most relevant terms in this limit,~\eqref{eq:mppp} gives
\begin{equation}
  \label{eq:M+++soft}
  \begin{split}
  \frac{\mathcal{M}^{\tilde{t}_i}_{+++}}{g_{h\, \tilde{t}_i \tilde{t}_i} v \Delta}& \simeq \frac{16}{tu}-\frac{16 m^2 }{m_H^2 t u}\left(t C_0(t)+u C_0(u)-2 m_H^2C_0 (m_H^2)\right)\\
  & -\frac{8 m^2}{m_H^2 t u}\left(st D_0(s,t)+su D_0(s,u)-tu D_0(t,u)\right)\,.
  \end{split}
\end{equation}
In the soft limit the relevant integral limits are
\begin{equation}
  \label{eq:soft-integrals}
  tC_0(t)\to 0\,,\quad uC_0(u)\to 0\,,\quad st D_0(s,t)\to 0\,,\quad us D_0(u,s)\to 0  \,,\quad ut D_0(u,t)\to 0\,,
\end{equation}
which gives
\begin{equation}
  \label{eq:M+++soft-fine}
  \begin{split}
 \mathcal{M}^{\tilde{t}_i}_{+++}  &\simeq \frac{16}{tu}{g_{h\, \tilde{t}_i \tilde{t}_i} v \Delta}\left(1+2 m^2 C_0 (m_H^2)\right)\\
  &\simeq (\sqrt{2})^3 \sqrt{\frac{s}{t u}} \mathcal{M}^{\tilde{t}_i}_{++}\,.
\end{split}
\end{equation}
Similarly, the other helicity amplitude~\eqref{eq:mppm} becomes
\begin{equation}
  \label{eq:M++-soft}
  \begin{split}
  \frac{\mathcal{M}^{\tilde{t}_i}_{++-}}{g_{h\, \tilde{t}_i \tilde{t}_i} v \Delta}& \simeq
  -\frac{16}{tu}+\frac{16 m^2}{m_H^2 tu}\left(t C_0(t)+u C_0(u)-2 m_H^2C_0 (m_H^2)\right)\\
    & -\frac{8 m^2}{m_H^2 tu}\left(st D_0(s,t)+su D_0(s,u)+tu D_0(t,u)\right)\,.
  \end{split}
\end{equation}
Evaluating again all scalar integrals in the soft limit we get
\begin{equation}
  \label{eq:M++-soft-fine}
  \begin{split}
 \mathcal{M}^{\tilde{t}_i}_{++-}  &\simeq -\frac{16}{tu}{g_{h\, \tilde{t}_i \tilde{t}_i} v \Delta}\left(1+2 m^2 C_0 (m_H^2)\right)\\
  &\simeq -(\sqrt{2})^3 \sqrt{\frac{s}{t u}} \mathcal{M}^{\tilde{t}_i}_{++}\,.
\end{split}
\end{equation}
These expressions have to be compared with the universal
behavior of helicity amplitudes \cite{Mangano:1990by}\cite{Dixon:1996wi}\footnote{The $\sqrt{2}$ factors comes from the differing normalisation factors for gauge group generators $\operatorname{tr}[T^a T^b] = \delta^{ab}$ in the spinor helicity formalism, compared to the usual $\operatorname{tr}[T^a T^b] = \frac{1}{2}\delta^{ab}$. This is compensated by a relative $\sqrt{2}$ factor associated to the gauge coupling.}:
\begin{equation}
  \label{eq:Msoft-universal}
  \begin{split}
    \mathcal{M}^{\tilde{t}_i}_{+++} & = (\sqrt{2})^3 \frac{\langle p_1 p_2\rangle}{\langle p_1 p_3
      \rangle \langle p_3 p_2 \rangle}  \mathcal{M}^{\tilde{t}_i}_{++}\,, \\
     \mathcal{M}^{\tilde{t}_i}_{++-} & = -(\sqrt{2})^3\frac{[ p_1 p_2]}{[ p_1 p_3] [ p_3 p_2]}  \mathcal{M}^{\tilde{t}_i}_{++}\,.
  \end{split}
\end{equation}
Since we have not used the spinor-helicity formalism, it is not
immediate to rephrase our expressions in terms of helicity
products. However, for real momenta, spinor products are simply equal to the square root of the relevant momentum invariant, up to a phase. The universal soft factor has an implicit helicity set by the helicity of the soft gluon, and so the choice of translating to angle or square bracket spinor products is fixed by this. We then obtain from ~\eqref{eq:M+++soft-fine} and~\eqref{eq:M++-soft-fine} that $\mathcal{M}^{\tilde{t}_i}_{+++}$ and $\mathcal{M}^{\tilde{t}_i}_{++-}$ have the
correct behavior~\eqref{eq:Msoft-universal} in the soft limit, modulo an overall phase that
depends on the gluon helicity. This phase is the same as for the
standard model case, and therefore can be factored out of each helicity
amplitude and will not contribute to the amplitude squared. 

\subsection{Collinear limits}

We consider the collinear limit $u\to 0$ where $p_1$ becomes collinear to $p_3$. Introducing the splitting fraction $z = \frac{m_H^2}{s}$, the invariants take the limiting values
\begin{equation}
  \label{eq:u->0}
  u\to 0\,,\quad s =\frac{m_H^2}{z}\,,\quad t\to -\frac{1-z}{z}m_H^2\,,\quad s_1\to -t\,,\quad t_1\to -s\,, \quad u_1 \to -m_H^2\,.
\end{equation}
In this limit $uC_0(u) \to 0$, whereas $sC_0(s)$ and $tC_0(t)$ stay finite. For the box integrals, we have
\begin{equation}
  \label{eq:Box-collinear}
  st D_0(s,t)\to 2\left[sC_0(s)+tC_0(t)-m_H^2 C_0(m_H^2)\right]\,,\quad suD_0(s,u)\to 0\,,\quad ut D_0(u,t)\to 0\,. 
\end{equation}
In this limit we get
\begin{equation}
  \label{eq:M+++collinear}
  \begin{split}
    \frac{ \mathcal{M}^{\tilde{t}_i}_{+++}}{g_{h\, \tilde{t}_i \tilde{t}_i} v \Delta}& \simeq \frac{16}{tu}\left(1+\frac{t}{u_1}\right)+\frac{32 m^2}{u u_1^2} u_1 C_1(u)\\ &  -\frac{16 m^2}{stu}\left[s_1 C_1(s)-\frac{s}{t_1}t_1 C_1(t)+\frac{t-s}{u_1} u_1 C_1(u)\right]\\
    & +\frac{8 m^2}{stu}\left[st D_0(s,t)+su D_0(s,u)-tu D_0(t,u)\right]\\
    & \simeq -\frac{16 z}{(1-z)u}\left(1+\frac{1-z}{z}\right)-\frac{32
      m^2}{m_H^4 u} m_H^2 C_0(m_H^2)\\ & + \frac{16 m^2 z^2}{(1-z)
      m_H^4 u} \left[sC_0(s)-m_H^2 C_0(m_H^2)+t C_0(t)-m_H^2
      C_0(m_H^2)-\frac{2-z}{z}m_H^2 C_0(m_H^2)\right]\\
    & -\frac{16 m^2 z^2}{(1-z)m_H^4 u}\left[sC_0(s)+t C_0(t)-m_H^2 C_0(m_H^2)\right] = -\frac{16}{(1-z)m_H^2 u}\left[1+2 m^2 C_0(m_H^2)\right]\,.
  \end{split}
\end{equation}
Similarly, for the other helicity configuration we obtain
\begin{equation}
  \label{eq:M++-collinear}
  \begin{split}
    \frac{ \mathcal{M}^{\tilde{t}_i}_{++-}}{g_{h\, \tilde{t}_i \tilde{t}_i} v \Delta}&\simeq -\frac{16 m_H^2}{stu}\left\{
      1-\frac{m^2}{m_H^2}\left[sC_0(s)-m_H^2 C_0(m_H^2)+t C_0(t)-m_H^2
      C_0(m_H^2)-m_H^2 C_0(m_H^2)\right]
    \right.\\
& \left. 
+ \frac{m^2}{m_H^2}\left[sC_0(s)+t C_0(t)-m_H^2 C_0(m_H^2)\right] \right\}\simeq \frac{16\, z^2}{(1-z)m_H^2 u} \left[1+2 m^2 C_0(m_H^2)\right]\,.
  \end{split}
\end{equation}

Now in the collinear case the limit depends on the helicity of each collinear leg. This means that there are two more possibilities to consider, and therefore we should additionally look at the limit of the two helicity
amplitudes $ \mathcal{M}^{\tilde{t}_i}_{-+-}$ and $ \mathcal{M}^{\tilde{t}_i}_{-++}$. The first can simply be found by interchanging $s$ and $t$ in $ \mathcal{M}^{\tilde{t}_i}_{+++}$. As this does not affect the relevance of terms in this limit, the switch can be effected by making the substitution $z \to \frac{z}{z-1}$ in the limit form, and so from~\eqref{eq:M+++collinear} we have
\begin{equation}
  \label{eq:M-+-collinear}
  \begin{split}
    \frac{ \mathcal{M}^{\tilde{t}_i}_{-+-}}{g_{h\, \tilde{t}_i \tilde{t}_i} v \Delta} &\simeq -\frac{16}{(1-\frac{z}{z-1})m_H^2 u}\left[1+2 m^2 C_0(m_H^2)\right]\\
    &\simeq-\frac{16}{m_H^2 u} (1-z) \left[1+2 m^2 C_0(m_H^2)\right]\,.
  \end{split}
\end{equation}
Extracting the collinear limit from $\mathcal{M}^{\tilde{t}_i}_{-++}$ is trickier. It is obtained from $\mathcal{M}^{\tilde{t}_i}_{+++}$ by exchanging $s$ and $u$, and as such the relevant terms in the collinear limit will be different in structure from the above cases. One has
\begin{equation}
  \label{eq:M-++collinear}
  \begin{split}
    \frac{\mathcal{M}^{\tilde{t}_i}_{-++}(s,t,u)}{g_{h\, \tilde{t}_i \tilde{t}_i} v \Delta} & =\frac{ \mathcal{M}^{\tilde{t}_i}_{+++}(u,t,s)}{g_{h\, \tilde{t}_i \tilde{t}_i}
      v \Delta}\\
    & \simeq \frac{16}{u}
    \left(\frac{t}{s_1^2}B_1(s)+\frac{s}{t_1^2}B_1(t)\right)\\ &
    - \frac{16 m^2}{s t u}\left(u_1 C_1(u)+\frac{s}{t_1} t_1
      C_1(t)+\frac{t}{s_1} s_1 C_1(s)\right)\\ & +\frac{8
      m^2}{stu}\left(utD_0(u,t)+us D_0(u,s)-st
      D_0(s,t)\right)-\frac{16 m^2}{stu} st
    D_0(s,t)+\frac{8}{u^2}E_0(s,t)\,.
  \end{split}
\end{equation}
Notice that, since the term containing $E_0(s,t)$ is proportional to
$1/u^2$, one needs to keep the linear terms in $u$ in the small-$u$
expansion of $E_0(s,t)$. In particular, as $E_0(s,t)$ is the linear combination defined in \eqref{eq:E0-definition} one cannot use the limit of
eq.~\eqref{eq:Box-collinear} to evaluate $stD_0(s,t)$, but rather one must use the extended version
\begin{align}
  \label{eq:stD-improved}
  stD_0(s,t)&\to2\left(1-\frac{2m^2 u}{st}\right)\left[sC_0(s)+tC_0(t)-m_H^2 C_0(m_H^2)\right]\nonumber\\
	&\quad+\frac{2u}{st}\left[sB_0(s)+tB_0(t)-m_H^2 B_0(m_H^2)\right]\,.
\end{align}
Substituting this expression in~\eqref{eq:M-++collinear} leads to
$\mathcal{M}_{-++}\simeq 0$ in the collinear limit $u\to 0$.

Collecting all results we have
\begin{equation}
  \label{eq:M-coll-all}
  \begin{split}
 \mathcal{M}^{\tilde{t}_i}_{+++} & \simeq  \frac{-(\sqrt{2})^3}{z\sqrt{(1-z)}\sqrt{-u}}\mathcal{M}^{\tilde{t}_i}_{++}\,,\\
 \mathcal{M}^{\tilde{t}_i}_{++-}&\simeq  \frac{z(\sqrt{2})^3}{\sqrt{(1-z)}\sqrt{-u}}\mathcal{M}^{\tilde{t}_i}_{++}\,, \\
 \mathcal{M}^{\tilde{t}_i}_{-+-} & \simeq
\frac{-(1-z)^2(\sqrt{2})^3}{z\sqrt{(1-z)}\sqrt{-u}}\mathcal{M}^{\tilde{t}_i}_{++}\,, \\
 \mathcal{M}^{\tilde{t}_i}_{-++}& \simeq 0\,.
\end{split}
\end{equation}
To check the correctness of the above limits, we have to translate our conventions for helicity and splitting fraction into the ones available in the literature, in which all momenta are considered to be outgoing. First, we need to swap the helicity of each incoming particle. Additionally, the relation of $z$ to the momenta is different when the collinear gluons are outgoing. One can switch between the two cases by making the replacement $z \to \frac{1}{z}$.  Adopting the usual convention of associating negative momentum signs to angle spinors we expect the behavior  \cite{Mangano:1990by,Dixon:1996wi}
\begin{equation}
 \label{eq:M-coll-universal}
 \begin{split}
\frac{\mathcal{M}^{\tilde{t}_i}_{+++}}{\mathcal{M}^{\tilde{t}_i}_{++}}& \simeq \operatorname{Split}_{+}\left(-1^-,3^+;\frac{1}{z}\right)= \frac{-(\sqrt{2})^3}{z\sqrt{1-z}\langle p_1
     p_3\rangle}\,, \\
\frac{\mathcal{M}^{\tilde{t}_i}_{++-}}{\mathcal{M}^{\tilde{t}_i}_{++}}&\simeq \operatorname{Split}_{+}\left(-1^-,3^-;\frac1z\right)= \frac{z(\sqrt{2})^3}{\sqrt{1-z}[p_1 p_3]} \,, \\
\frac{\mathcal{M}^{\tilde{t}_i}_{-+-}}{\mathcal{M}^{\tilde{t}_i}_{++}}& \simeq \operatorname{Split}_{+}\left(-1^+,3^-;\frac 1z\right)= \frac{-(1-z)^2(\sqrt{2})^3}{z\sqrt{1-z}\langle p_1
     p_3\rangle}\,,\\
\frac{\mathcal{M}^{\tilde{t}_i}_{-++}}{\mathcal{M}^{\tilde{t}_i}_{++}}& \simeq \operatorname{Split}_{+}\left(-1^+,3^+;\frac 1z\right) = 0\,.
\end{split}
\end{equation}

We must now translate~\eqref{eq:M-coll-all} to helicity language. The translation from Mandelstam variables to spinor invariants is similar to the soft case, although the helicity consideration is slightly subtler. As the three legs of the splitting amplitude are collinear, we no longer have information about the contribution from each individual leg, as the helicity spinors become proportional. Instead what matters is the overall (outgoing) helicity of the three, which governs whether it is appropriate to translate to angle or square brackets, and with this consideration we indeed find the correct momentum dependence. However, this is not relevant in the end because, up to an overall phase $[p_1 p_3]\sim \langle p_1 p_3\rangle\sim \sqrt{-u} $.

\end{document}